\documentclass[a4paper,11pt]{article}
\usepackage{jinstpub} 
\usepackage{lineno}
\usepackage{float}
\usepackage{subfigure}
\usepackage{caption}
\usepackage{amsmath, siunitx}
\usepackage{lipsum}
\usepackage{cite}
\bibpunct{[}{]}{,}{n}{,}{,}
\usepackage[colorlinks=true, linkcolor=blue]{hyperref}

\title{\boldmath Unified framework for precise background modeling to enhance rare event detection at the Kuo-Sheng reactor-neutrino laboratory}

\author[a,1]{Subhasis Parhi,\note{Corresponding author.}}
\author[a]{Lakhwinder Singh,}
\author[b]{Manoj Kumar Singh,}
\author[b]{Henry Tsz-King Wong,}
\author[a,2]{and Venktesh Singh,\note{Corresponding author.}}

\affiliation[a]{Department of Physics, Central University of South Bihar, Gaya 824236, India.}
\affiliation[b]{Institute of Physics, Academia Sinica, Taipei 115201, Taiwan.}

\emailAdd{subhasisparhi.official@gmail.com; venktesh@cusb.ac.in}

\abstract{A comprehensive GEANT4 simulation framework is
  developed to model the background of the TEXONO
  experiment, considering contributions from radioactive
  isotopes in both the detector components and the
  ambient environment. In this framework, the HPGe
  detector's front-end electronics (pre-amplifier) are
  modeled to contain trace amounts of naturally occurring
  radionuclides $^{238}$U, $^{232}$Th, and $^{235}$U from
  manufacturing materials. The results confirm that the
  decay chains of $^{238}$U and $^{232}$Th dominate the
  background in this region. The observed background
  contributions from both isotopes are at the
  $\mathcal{O}$(1)~counts~kg$^{-1}$keV$^{-1}$day$^{-1}$ for
  energies below 400~keV. Trace radioisotopic impurities
  were introduced into the anti-Compton veto (ACV)
  detectors to reflect realistic material compositions:
  $^{40}$K was incorporated in the NaI(Tl) crystal consistent
  with its natural isotopic abundance, and $^{137}$Cs was
  included in the CsI(Tl) detector, representing a
  common anthropogenic contaminant. The analysis
  identifies minor yet measurable background components
  originating from $^{40}$K in the NaI(Tl) ACV detector
  and $^{137}$Cs in the CsI(Tl) scintillator. The
  residual spectrum is dominated by $^{40}$K
  $\gamma$-rays, with smaller components from
  $^{137}$Cs, in agreement with simulation results. The
  background rate from $^{40}$K is uniform at
  $\sim$0.1~counts~kg$^{-1}$keV$^{-1}$day$^{-1}$,
  approximately 10 times higher than that from
  $^{137}$Cs below 400~keV. To incorporate environmental
  sources of radioactivity, simulations included
  isotopes $^{60}$Co, $^{54}$Mn, and $^{135}$Xe
  distributed within the air gap between the
  copper end-cap and the NaI(Tl) ACV detector,
  representing airborne and surface-induced
  contamination. Although these sources contribute
  to the overall background (10$^{-2}$, 10$^{-2}$,
  and 0.1~counts~kg$^{-1}$keV$^{-1}$day$^{-1}$ below
  100~keV for $^{135}$Xe, $^{54}$Mn, and  $^{60}$Co,
  respectively), their effects are well controlled
  and constitute only minor components of the total
  background. The comparison between simulated and
  measured spectra, while showing minor deviations
  at specific $\gamma$-lines, demonstrates the
  validity of the background model and the robustness
  of the simulation framework for guiding detector
  and shielding design.
}


\arxivnumber{2510.05868}

\begin{document}
\maketitle
\flushbottom

\section{Introduction}
\label{sec:intro}
Searches for rare observables such as dark matter (DM)~\citep{CDEX::PRL:2018,Li::2025},
coherent elastic neutrino-nucleus scattering (CE$\nu$NS)~\citep{Karmakar::2025,Wong::2025,WongHT::2006,Akimov::2017},
neutrinoless double-beta decay (0$\nu\beta\beta$)~\citep{Singh::2020,Singh::2024,Singh::2025:IJMPA}, and
sterile neutrinos~\citep{BD::JK::2021}, etc. require ultra-low
background conditions, as background events can dominate the
measured spectrum and severely limit the experiment's discovery
potential. Significant experimental advancements have been realized
over the past several decades across various energy regimes,
depending on the physics objectives. Yet, the pursuit of rare-event
phenomena continues to be limited by background and detector
constraints, highlighting the need for ongoing innovation in
material purity, shielding design, and signal discrimination
techniques~\citep{Aalseth::2013,Ackerman::5206,Agostini::2020,
  Aprile::2018}. In recent years, discussions on zero-background and
no-threshold detection systems emphasize their potential to
significantly enhance interaction sensitivity and allow the use
of more compact detector designs. Although the complete
elimination of background sources is practically impossible,
their effective suppression remains a critical requirement for
improving detector performance and experimental
sensitivity~\citep{Dokania::2014,Leonard::2008,Arnold::1995}.

The TEXONO experiment ({\bf T}aiwan {\bf EX}periment {\bf O}n {\bf N}eutrin{\bf O})~\citep{Wong::2007}
represents one of the foremost international collaborations
dedicated to probing rare low-energy physics processes. Its
research program spans a wide range of topics, including
precision studies of reactor neutrinos through
neutrino-electron scattering, searches for neutrino
electromagnetic properties such as magnetic moment and
millicharge, CE$\nu$NS, and DM interactions, encompassing
both spin-independent and spin-dependent WIMPs, bosonic
super-WIMPs, axions, and fractionally charged
particles~\citep{HTW::Uni::2015}. The TEXONO Collaboration
is a founding partner and active participant of the CDEX
(China Dark matter Experiment) Collaboration. The long-term
objective of the joint TEXONO-CDEX program is to develop a
ton-scale high-purity germanium detector array (CDEX-1T and
CDEX-10T) at the China Jinping Underground Laboratory (CJPL)
for the searches of DM and 0$\nu\beta\beta$. At present,
the collaboration employs kilogram-scale point-contact
high-purity germanium (PCGe) detectors with multiple readout
channels, enabling sensitivity across a wide energy range
from sub-keV thresholds (below 500~eV) for DM searches to
high-energy regions (up to 3000~keV) relevant for
0$\nu\beta\beta$ investigations~\citep{CDEX::Ge::2017}.

\begin{figure}[h!]
   \centering 
   \includegraphics[width=0.8\textwidth]{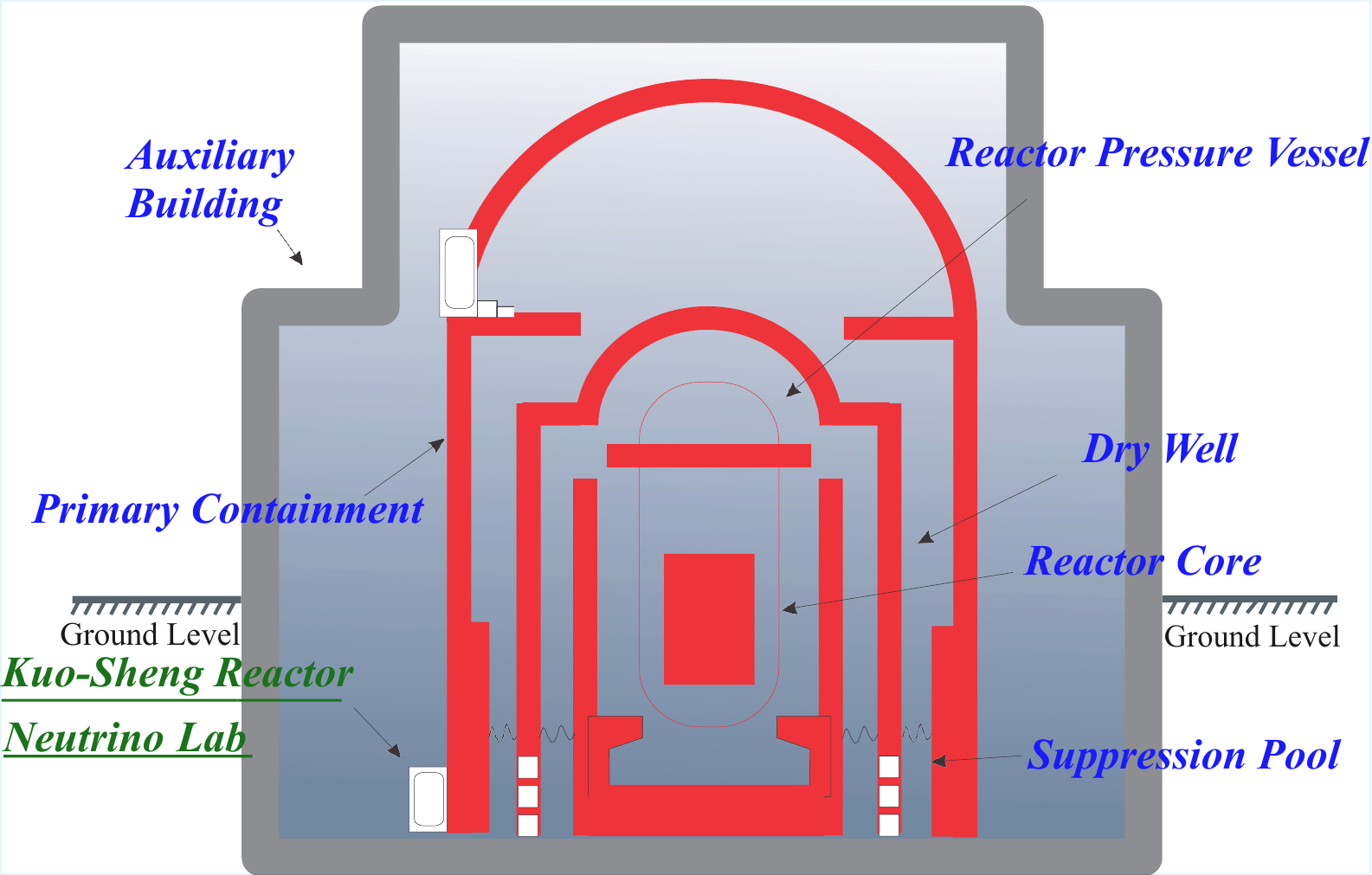}	
   \caption{A schematic side view (not drawn to scale) of the
     Kuo-Sheng Nuclear Power Station Reactor Building is
     presented, indicating the placement of the experimental
     site relative to the reactor core. The detector is
     positioned at an approximate core-to-detector distance of
     28~m, which defines the experimental baseline for
     low-energy neutrino and rare-event measurements. Reprinted (figure) with permission from ~\citep{Wong::2007}, Copyright
(2007) by the American Physical Society.} 
   \label{fig::site::ksnl}
 \end{figure}

In such rare-event physics searches, achieving high
sensitivity and reliable signal identification requires a
comprehensive understanding and precise modeling of all
background contributions arising from both
detector-intrinsic and environmental sources. Careful
characterization and suppression of these backgrounds are
essential to fully exploit the discovery potential of the
TEXONO-CDEX program. In this direction, a detailed
understanding of the nature and sources of background
events, as well as the implementation of robust reduction
and rejection techniques, is indispensable for improving
the sensitivity and reliability of rare-event
measurements. In order to achieve a precise understanding
of background contributions, the present work considers
the TEXONO experimental setup at the Kuo-Sheng reactor-neutrino
laboratory (KSNL). A dedicated background model
is developed using Monte Carlo (MC) simulations that
account for the full range of environmental radiation
sources and experimental components, providing a realistic
framework for the physics analysis.

To establish a comprehensive understanding of the
detector's background, this study employs data acquired from
a coaxial-type HPGe detector, as described in
Section~\ref{sec::det}. This approach facilitates a robust
characterization of the overall background environment
essential for rare-event analysis. The total background
within the active volume of the
detector can be broadly classified into two categories:
(a) internal backgrounds, primarily due to cosmogenic
activation, and (b) external backgrounds arising from
photons, neutrons, and cosmic-ray muons. 
Particles produced by cosmic-ray muon interactions and
$\gamma$-rays emitted from the decay of naturally
occurring radioactive nuclides constitute significant
sources of external background, from which detectors
must be effectively shielded, which is a challenge
that is particularly critical for high-purity germanium
(HPGe) detectors. Cosmic-ray muon contributions are
controlled by employing a dedicated muon-veto system
that achieves a $\sim$92-93\% rejection
efficiency~\citep{MSingh::CJP2019},
notwithstanding the low overburden of approximately
30~m.w.e. (meter water equivalent) at the site. Therefore,
in the primary TEXONO background simulation,
muon-induced and cosmogenic components are not included,
as these contributions are effectively handled at the
analysis stage through the active muon-veto system and
event-selection criteria. At the KSNL, the dominant sources of
environmental radioactivity arise from the presence of
$^{238}$U, $^{232}$Th, and $^{40}$K in the surrounding
construction materials, detector supports, and shielding
components. To maintain clarity in interpretation, the
background model developed
in this work includes only external radiation originating
from natural radioisotopes. Specifically, the contributions
from the $^{238}$U and $^{232}$Th decay chains arise from
the front-end electronics (pre-amplifier) of the HPGe
detector, while $^{137}$Cs is associated with the CsI(Tl)
detector and $^{40}$K originates from the NaI(Tl)
detector used as the active Anti-Compton Veto (ACV).
Furthermore, the model also includes contributions from 
$^{60}$Co, $^{54}$Mn, and $^{135}$Xe, which serve as
additional photon background sources resulting from
reactor-related activation and may be present in airborne
dust near the laboratory.

MC simulations are employed to
reproduce the detector response and evaluate the relative
contributions from each background component, providing a
realistic framework for quantitative comparison with the
measured TEXONO data. We emphasize that the present
framework does not constitute a fully predictive model,
rather, it is a data-constrained approach in which model
components are tuned and validated against measured
spectra to enable process-resolved background decomposition.
Looking forward, we plan to extend the simulation
framework to the case of $p$-type PCGe detectors. The
dominant difference between
these detector setups lies in their geometrical configurations,
which govern their respective sensitivities to
surface-related backgrounds. While surface $\alpha$
backgrounds are not explicitly simulated in the present
work, their contribution can be effectively mitigated
through pulse-shape based surface event rejection
techniques~\citep{HBL::Astro2014,LTY::NIMA2018}.
By adapting the simulation to the PCGe configuration,
we aim to achieve a quantitative and process-resolved
understanding of the background contributions present
in the current dataset, particularly those
intrinsically localized within the shielding volume.
In this framework, the simulation enables {\it a priori}
estimation of internally generated backgrounds,
apart from cosmogenic activation, thereby supporting
predictive, simulation-assisted subtraction
strategies~\citep{Ackerman::5206} rather than reliance
on shielding optimization, which is not feasible for
these sources. An important exception is the
reactor-associated gaseous $^{135}$Xe component, whose
contribution is mitigated operationally through improved
and continuous purging with nitrogen gas derived from
liquid-nitrogen boil-off~\citep{Wong::2007},
as implemented in the present setup. Collectively,
these approaches offer guidance for material radiopurity
control and related mitigation pathways in ongoing
and future measurements, although they lie beyond the
scope of the present study and are not accompanied by
quantitative estimates here.

Despite variations in contamination levels and operating
conditions, this work demonstrates that, given a
well-defined geometry, material inventory, and screening
inputs, MC simulations can reproduce experimental
backgrounds with good fidelity, thereby enabling a
data-constrained modeling approach that extends beyond
purely empirical background subtraction in rare-event searches,
where backgrounds must be reliably controlled prior to
construction and data taking. With the Kuo-Sheng Nuclear
Power Station entering decommissioning and the experiment
planned for relocation under the CDEX collaboration to
the Sanmen Nuclear Power Station~\citep{Karmakar::2025},
the background modeling framework developed in this work
is not directly transferable, as site-dependent activity
concentrations require location-specific validation and
revision. Nevertheless, the framework provides a
structured and systematic basis that can be re-optimized
and re-tuned with site-specific inputs and measurements,
serving as a valuable and transferable tool for background
decomposition, simulation tuning, and for anticipating
as well as interpreting background conditions at the
relocated facility. It further alleviates reliance on
limited reactor-OFF data and enables robust background
subtraction from reactor-ON measurements, thereby
improving experimental sensitivity beyond purely data-driven
approaches.

\section{Experimental framework and methodological considerations}
The TEXONO Collaboration conducts a research program on
low-energy neutrino and DM physics at the KSNL, located
on the northern shore of Taiwan. The laboratory is
situated approximately 28~m from Core-1 of the Kuo-Sheng
Nuclear Power Station, operated by the Taiwan Power
Company, which has a nominal thermal power output of
2.9~GW. The relative location and proximity to the
reactor core are shown schematically in
figure~\ref{fig::site::ksnl}. The experimental site has an
overburden of about 30~m.w.e. (meter water
equivalent)~\citep{HTW::Uni::2015,Sing::2017,WongHT::2006,
  Soma::2016}. A brief account of
the experimental configuration is presented below, covering
the germanium detector, the active and passive shielding
arrangements, the data acquired, and the methodological
considerations applied in the present analysis.

\begin{figure}[h!]
   \centering 
   \includegraphics[width=0.8\textwidth]{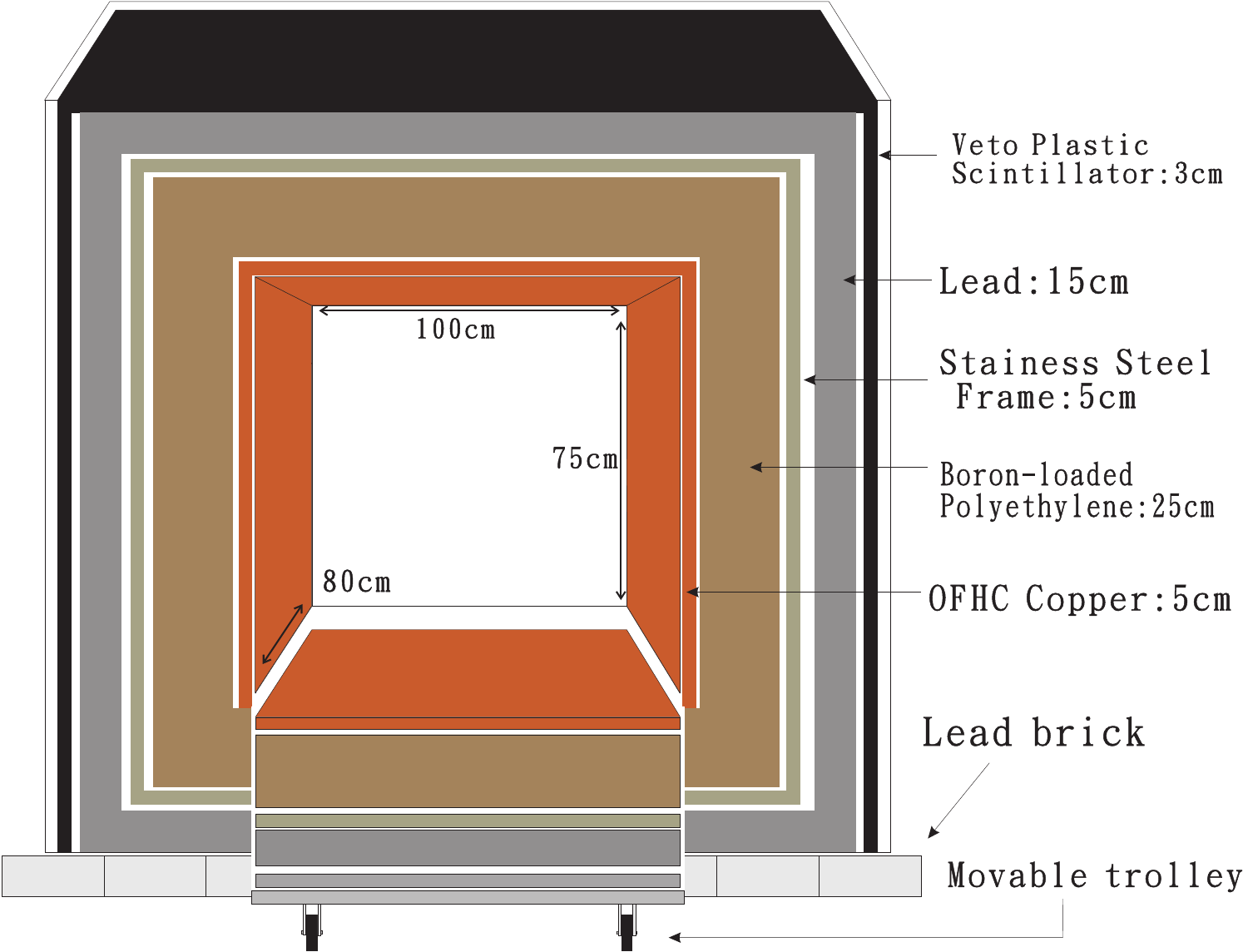}	
   \caption{A multi-purpose inner target volume is
     enclosed by layers of passive shielding composed
     of materials optimized to attenuate $\gamma$-rays
     and neutrons, complemented by cosmic-ray veto
     panels to identify and reject events induced by
     penetrating muons. Reprinted (figure) with permission from ~\citep{Wong::2007}, Copyright
(2007) by the American Physical Society.} 
   \label{inner::shield}
 \end{figure}

\subsection{Experimental setup at KSNL}
The laboratory is enclosed within a 50-ton shielding
assembly, which, proceeding from the exterior inward,
comprises 2.5~cm thick plastic scintillator panels on
five sides for cosmic-ray veto (CRV), followed by 15~cm
thick lead walls designed to attenuate high-energy
$\gamma$-rays~\citep{Wong::2004}. The second layer of
the passive shielding consists of 5~cm thick stainless
steel, which serves a dual purpose: providing
mechanical support for the entire shielding assembly
and reducing the $\gamma$-ray flux. Stainless steel,
primarily composed of iron, is effective in moderating
fast neutrons through inelastic scattering and also
acts as an efficient absorber of thermal
neutrons~\citep{ASonay::PRC2018}. The
subsequent shielding layer is 25~cm thick boron-loaded
polyethylene, which significantly attenuates the
incoming neutron flux from the environment. Finally, a
5~cm thick layer of oxygen-free high-conductivity (OFHC)
copper is employed to absorb photons originating from
intrinsic radioactive contaminants as well as
$\gamma$-rays produced by neutron or cosmic-ray
interactions in the lead and boron-loaded polyethylene
layers, as depicted in figure~\ref{inner::shield}. The
innermost volume, with dimensions of
100$\times$80$\times$75~cm$^{3}$, offers versatile
accommodation for the placement of different detector
types, thereby facilitating the exploration of diverse
physics objectives within the same experimental setup.

\begin{figure}[h!]
   \centering 
   \includegraphics[width=0.9\textwidth]{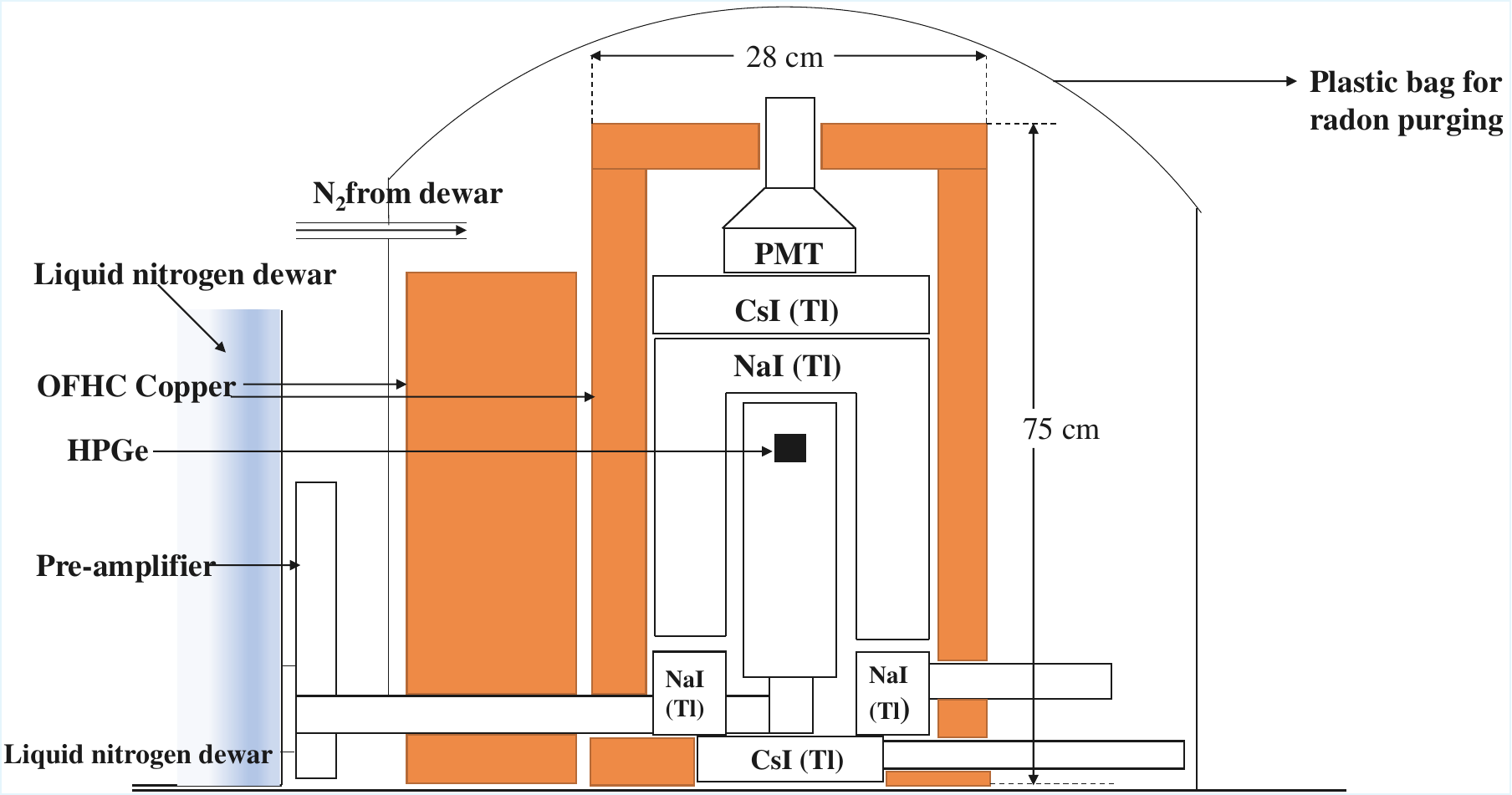}	
   \caption{Schematic diagram of the detector configuration
     used in this study, including the coaxial HPGe detector,
     surrounding anti-Compton veto (ACV) detectors, inner OFHC
     copper shielding, and the radon-purge system connected
     to the LN$_{2}$ dewar exhaust. Reprinted (figure) with permission from ~\citep{Wong::2007}, Copyright
(2007) by the American Physical Society.} 
   \label{fig::Schem::Setup}
 \end{figure}

The detection systems at the KSNL have evolved,
leveraging the superior performance of HPGe
detectors for low-energy physics. Earlier
experimental phases relied on coaxial HPGe detectors
with an active mass of around 1~kg, primarily used for
neutrino-electron scattering searches and setting
limits on the neutrino magnetic moment with a typical
energy threshold $\sim$5~keV~\citep{Wong::2007}.
Building upon its earlier efforts, the collaboration
advanced its research program toward the sensitivity
regime necessary for CE$\nu$NS and light DM
investigations by adopting next-generation PCGe
detectors~\citep{Karmakar::2025,LSingh::PRD2019,MSingh::CJP2019}.
The progression in detector technology plays an
essential role in extending the reach of low-energy
rare-event searches. One of the central long-term
goals is the realization of detectors capable of
sub-100-eV energy thresholds, extremely low intrinsic
backgrounds, and scalable masses at the kilogram
level, thereby enabling high-precision CE$\nu$NS
studies. In parallel, the TEXONO collaboration is
developing advanced software frameworks and analysis
algorithms aimed at enhancing offline background
suppression and reducing the effective analysis
threshold~\citep{JSW:EPJC:2025}, thereby improving
the overall sensitivity of the experiment.

\subsection{Measured data selection and background modeling considerations}
\label{sec::det}
The detector utilized in this work is a coaxial-type HPGe
detector with an active mass of 1.06~kg~\citep{Wong::2007,
  Canbera::5019}. It features a Li$^{+}$-diffused outer
electrode of 0.7~mm thickness, and all detector components
and housing materials are selected to satisfy
ultra-low-background (ULB) design requirements. The end-cap
cryostat, also 0.7~mm thick, is fabricated from OFHC copper.
This design
effectively suppresses ambient $\gamma$-background below 60~keV,
ensuring that events in this region originate primarily from
MeV $\gamma$-rays via internal activity or Compton scattering.
As a result, the background spectrum below 60~keV remains
smooth and continuous, a feature essential for precise
studies of nuclear effects. The HPGe detector is enclosed
within a 4$\pi$ anti-Compton veto (ACV) system consisting of:
(1) a 5~cm thick NaI(Tl) ``well detector'' mounted on the
end-cap cryostat and optically coupled to a 12~cm PMT
through a 7~cm CsI(Tl) crystal serving as an active light
guide; (2) a 5~cm thick NaI(Tl) ``ring detector'' positioned
around the cryostat joint; and (3) a 4~cm thick CsI(Tl)
``base detector'' placed beneath the assembly. All ACV
detectors are supported by OFHC copper structures and read out by
photomultiplier tubes (PMTs) fabricated from low-radioactivity
glass. The entire assembly is further enclosed by an inner
shield consisting of 3.7~cm thick OFHC copper, with an
additional 10~cm OFHC copper wall providing lateral shielding
for the LN$_{2}$ dewar and preamplifier electronics. The inner
shielding and detector assembly are enclosed within a plastic
housing connected to the LN$_{2}$ dewar exhaust line, which
serves as a radon-purging system to minimize the accumulation
of radioactive radon gas. A schematic representation of the
detector configuration employed in this study at KSNL is
presented in figure~\ref{fig::Schem::Setup} .

\begin{figure}[h!]
  \centering 
  \includegraphics[width=\textwidth]{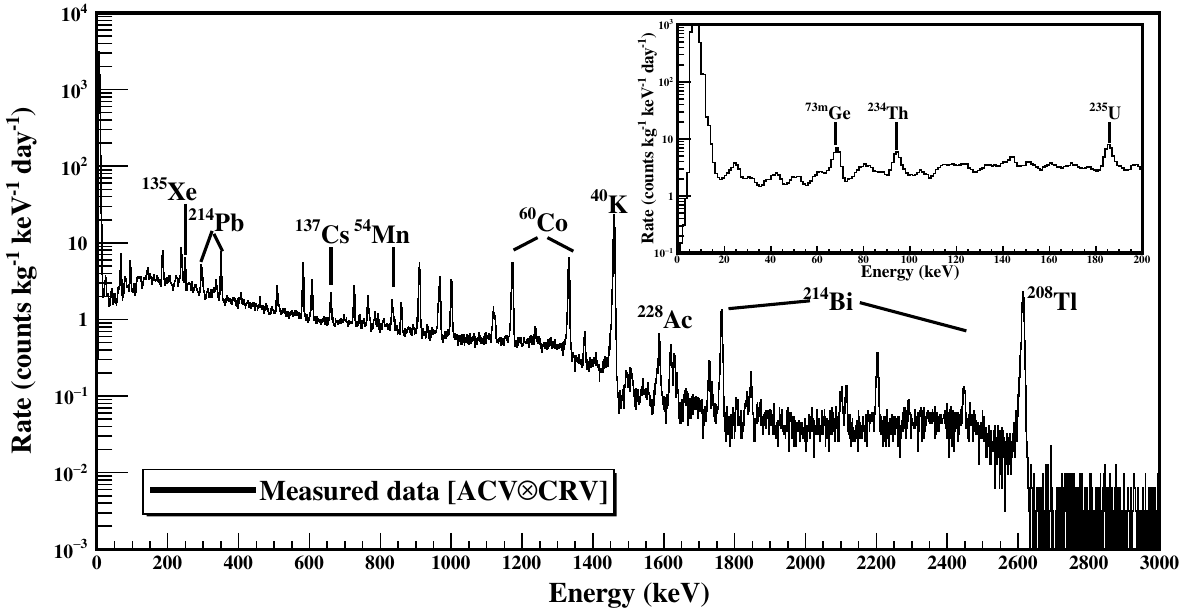}	
  \caption{The uncorrelated (ACV$\otimes$CRV) spectrum,
    constructed after applying the complete set of selection
    criteria, is utilized as the reference for this study
    across the full energy range~\citep{Wong::2007}. The inset
    highlights the sub-200~keV region, where the major
    $\gamma$-lines are well resolved and identified for
    calibration and background characterization purposes. Reprinted (figure) with permission from ~\citep{Wong::2007}, Copyright
(2007) by the American Physical Society.} 
  \label{fig::data}
\end{figure}

In the present study, uncorrelated (ACV$\otimes$CRV) data
tagged only with the germanium detector are used as a reference
for background characterization over the energy range from
threshold up to 3~MeV. A detector threshold of 5~keV and a
background level of approximately 1 event~keV$^{-1}$kg$^{-1}$day$^{-1}$
above 12~keV were achieved. The spectrum corresponding to
Period-III (taken from Ref.~\citep{Wong::2007}) is shown in
figure~\ref{fig::data}, where all identified $\gamma$-lines
are indicated. The spectrum represents data after applying
the full set of analysis efficiency and correction procedures.
In the present work, GEANT4 MC simulations are performed
for background modeling using the detector and shielding
components described above as input. The measured spectrum
shown in figure~\ref{fig::data} serves as the reference for
comparison with the simulated results. It is important to
note that the 66.7~keV $^{73m}$Ge line observed in the
experimental spectrum originates from cosmogenic activation,
which is not incorporated in our GEANT4 background model. For
completeness, we retain this peak in the data displayed in
figure~\ref{fig::data}, its absence in the simulation is
therefore expected and does not impact the optimization of
the modeled background components. A detailed description
of the simulation inputs and the corresponding comparative
analysis is presented in the following sections.

\begin{figure}[h!]
  \centering 
  \includegraphics[width=1.0\textwidth]{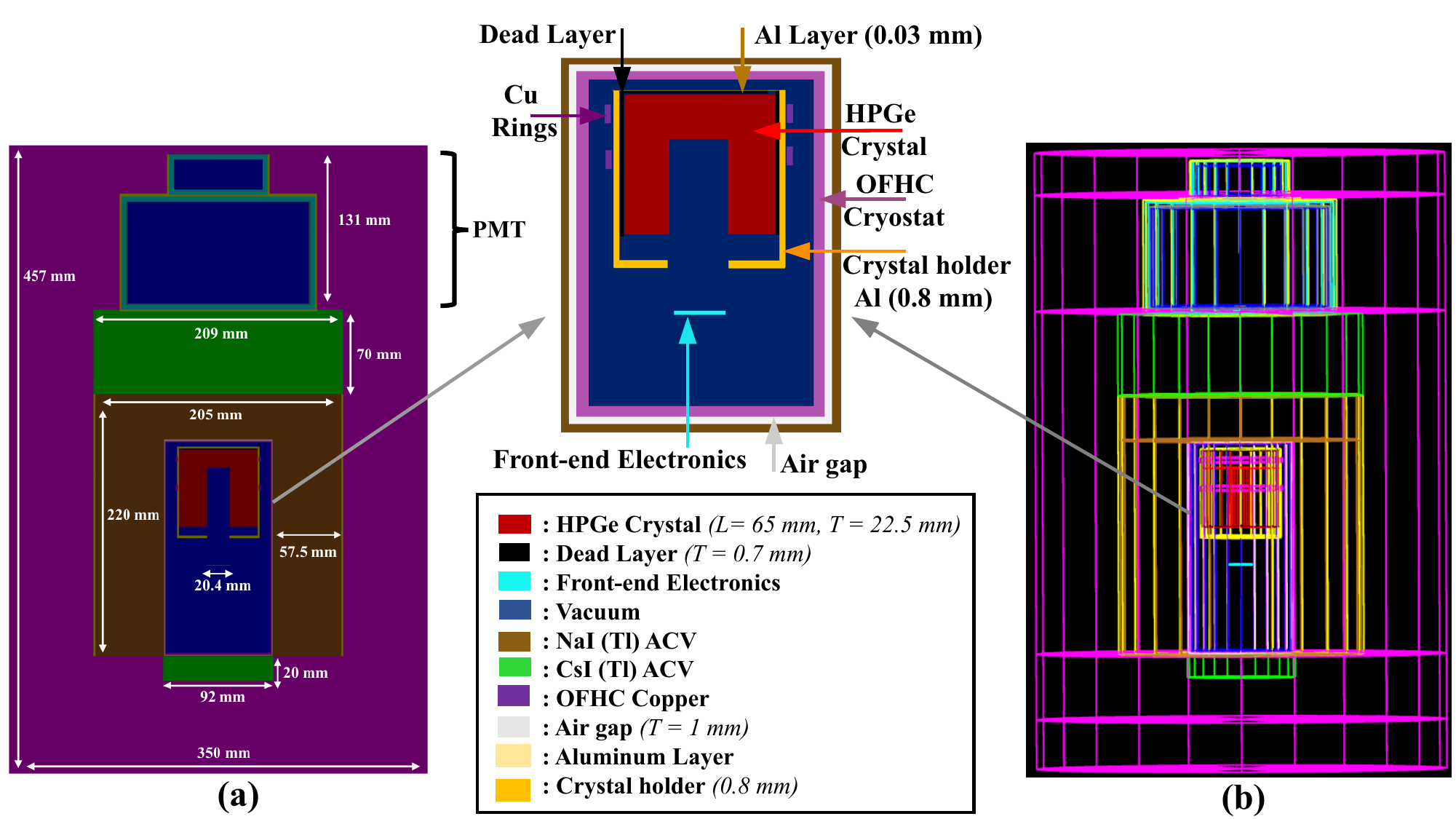} 
  \caption{(a) Semicircular cross-sectional representation of
    the detector geometry and associated active shielding
    used in the present background modeling. The schematic
    indicates relevant dimensional parameters, where $L$
    denotes the length and $T$ the thickness of the corresponding
    components. (b) A schematic of the entire detector simulation,
    including peripheral and auxiliary elements, is presented
    to illustrate the geometrical implementation in the
    model. The legend box explicitly illustrates the detector
    components together with their color coding and the relevant
    component dimensions utilized in the simulation. The inset
    at the center clearly depicts the SCGe detector
    together with the immediately surrounding components associated
    with it.} 
  \label{fig::Setup::Used}
\end{figure}

\section{GEANT4-based simulation framework for detector response modeling}
A Monte Carlo-based background model has been developed for
the TEXONO experiment at KSNL using the GEANT4 simulation
toolkit~\citep{geant4::2003}. Designed for detailed modeling
of particle-matter interactions, GEANT4 allows consistent
treatment of electromagnetic, hadronic, and radioactive
decay processes across energy scales. The implemented
simulation includes all relevant physical processes and
corresponding cross sections through dedicated data
libraries. The simulation accounts for all contributions
from the experimental setup and surrounding environment, with
the dimensions of the shielding, detectors, and associated
components precisely matching those of the KSNL configuration.

A detailed semicircular cross-sectional diagram of the
simulated detector assembly, including both the germanium
detector and its active shielding components, is shown in
figure~\ref{fig::Setup::Used}(a). This configuration forms the
basis of the present background modeling. It shows the
fabricated structure of the segmented coaxial germanium
(SCGe) detector, along with detailed component specifications
and dimensions as implemented in the simulation model.
Although Ref.~\citep{Wong::2007} does not provide precise
specifications of the crystal holder, we have incorporated
its dimensions in our simulation based on the
manufacturer-provided data from Canberra (Mirion
Technologies)~\citep{Cryst:Hold:Canb}, complemented by
additional details available in
Refs.~\citep{Cryst:Hold:2025,Cryst:Hold:2023:HLee,
  Cryst:Hold:2023:MTrav}, in order to
achieve a realistic and accurate geometrical model for our
analysis. All dimensions are indicated in millimeters (mm). The
germanium detector is enclosed within a vacuum-sealed
cylindrical copper (Cu) housing and further surrounded by
NaI(Tl) and CsI(Tl) scintillator detectors, which
function as active veto layers to minimize the ambient
$\gamma$-background. Coincidences between the SCGe and the
NaI(Tl)/CsI(Tl) veto detectors are extracted and analyzed
using their respective energy spectra to characterize and
reject background events. Signal readout from the NaI(Tl)
detector is achieved via a coupled PMT. The detector and
all associated components are additionally enclosed within
a copper shielding structure to reduce ambient background.
The complete detector assembly, along with auxiliary edges
and supporting components, is shown in
figure~\ref{fig::Setup::Used}(b), providing a visualization of
the configuration accurately implemented in the simulation.

\subsection{Detector response folding and normalization of simulated event rates}
To accurately reproduce the detector response, the
simulated energy depositions were first convolved
with an energy-dependent resolution function prior
to constructing the final spectra. The intrinsic energy
resolution of the HPGe detector was parameterized using
calibration $\gamma$-ray peaks as
\begin{equation}
  \sigma_{E}(E) = \alpha\sqrt{E} + \beta, ~~\alpha = 4.36338\times10^{-2}, \beta = 4.31432\times10^{-1},
  \label{reso}
\end{equation}
where $E$ is the deposited energy in units of keV and
$\sigma_{E}$ is given in keV. The coefficients $\alpha$ and
$\beta$ were obtained by fitting the resolution function
to several $\gamma$-ray calibration lines (e.g. from
$^{135}$Xe, $^{137}$Cs, $^{60}$Co, etc.), ensuring that the adopted
parameterization reflects the experimentally measured
detector response. This empirical parameterization
reproduces the observed calibration peak widths and
gives $\sigma_{E}$(2.23~MeV)$\approx$2.49~keV
(FWHM$\simeq$5.9~keV), which agrees at the
$\approx$8\% level with the RMS resolution of 2.3~keV
at 2.23~MeV reported in Ref.~\citep{texono::axion::PRD2007}.
Each simulated event energy deposition was subsequently
smeared by convolution with a Gaussian kernel of standard
deviation $\sigma_{E}(E)$ prior to histogramming, such
that the resulting simulated spectrum accurately reflects
the finite resolution response of the HPGe detector
across the full energy range. The energy dependent
resolution function is uniformly applied across the
entire simulated spectrum, rather than optimized for
each peak individually. This ensures a consistent
treatment of all features and prevents artificial
deformation of relative peak intensities and
misidentification of overlapped transitions.

In the present simulation
framework, the generated spectra are recorded in counts,
ensuring that each energy bin directly corresponds to
the expected number of events for subsequent analysis.
After accounting for the detector's finite energy
response through the resolution smearing procedure,
to facilitate direct comparison with experimental
measurements, the simulated spectrum is presented in
terms of a normalized rate, with units of
counts~kg$^{-1}$keV$^{-1}$day$^{-1}$. The transformation
of simulated counts into a normalized spectrum considers
the activity of the isotope in the specific volume
($A_{\rm vol}$ in Bq/kg), the mass of that volume ($m_{\rm vol}$ in kg),
the mass of the HPGe detector ($m_{\rm det}$ in kg), and
the energy bin width ($\Delta E$ in keV).
The mathematical formalism for this conversion is
based on the computation of the number of decays per
day ($N_{\rm decay/day}$) arising from the $A_{\rm vol}$,
as detailed in the following expression
\begin{equation}
  N_{\rm decay/day} {=} A_{\rm vol} \cdot  m_{\rm vol} \cdot 86400, 
  \label{activity::eq}
\end{equation}
where 86400 represents the total number of seconds in a
day. This relation converts the activity per unit mass into
the total $N_{\rm decay/day}$ for the volume in which the
decay chain or isotope is confined. The corresponding
normalized rate in counts~kg$^{-1}$keV$^{-1}$day$^{-1}$ is
then obtained as
\begin{equation}
  \begin{split}
    R &= \frac{\rm{counts} \cdot N_{\rm decay/day}}{m_{\rm det}\cdot \Delta E}, \\
      &= \frac{\rm{counts} \cdot A_{\rm vol}\cdot  m_{\rm vol} \cdot 86400}{m_{\rm det}\cdot \Delta E}.
  \end{split}
  \label{rate::eq}
\end{equation}
This formulation provides a normalized spectrum that is
directly comparable to the measured data.

\section{Radioactive contamination in detector components}
Environmental radioactivity and cosmic-ray induced photons
constitute the two primary sources of photon background in
the experimental data. Environmental radioactivity arises
predominantly from naturally occurring long-lived isotopes
in the decay chains of uranium ($^{238}$U), thorium
($^{232}$Th), and potassium ($^{40}$K), both at KSNL and in
typical laboratory environments. The reference background
data measured at KSNL reveal the presence of approximately
fifteen radioactive isotopes, including Thallium
($^{208}$Tl), Bismuth ($^{214}$Bi and $^{212}$Bi), Actinium
($^{228}$Ac), Potassium ($^{40}$K), Cobalt ($^{60}$Co),
Protactinium ($^{234m}$Pa), Manganese ($^{54}$Mn), Lead
($^{214}$Pb and $^{212}$Pb), Caesium ($^{137}$Cs), Radium
($^{226}$Ra), Uranium ($^{235}$U), Thorium ($^{234}$Th), and
Germanium ($^{73m}$Ge). A previously unassigned 249~keV
$\gamma$-line was subsequently attributed to xenon
($^{135}$Xe).

\setlength{\tabcolsep}{2.11em}
\begin{table}[h!]
  \centering
  \caption{In this work, the $^{238}$U and $^{232}$Th decay
    series are treated as multiple sub-chains, and the
    activities of the respective isotopic groups are
    incorporated into the background model to accurately
    simulate their contributions.}
  \renewcommand{\arraystretch}{1.2}
	\begin{tabular}{c|c|c}
		\hline
  \textbf{Mother chains} & \textbf{Split chains} & \textbf{Activity (Bq/kg)}\\
  \hline \hline
		& $^{238}$U & 1.04 $\times$ 10$^{-5}$ \\
		\cline{2-3} 
		& $^{234}$Th to $^{234}$Pa & 1.75 $\times$ 10$^{-4}$ \\
		\cline{2-3} 
		& $^{234}$U & 5.8 $\times$ 10$^{-6}$\\
		\cline{2-3} 
		& $^{230}$Th & 1.0 $\times$ 10$^{-7}$\\
		\cline{2-3} 
 $^{238}$U       & $^{226}$Ra & 2.27 $\times$ 10$^{-5}$\\
		\cline{2-3} 
		& $^{222}$Rn to $^{210}$Tl & 1.54 $\times$ 10$^{-4}$\\
		\cline{2-3} 
		& $^{210}$Pb & 1.67 $\times$ 10$^{-8}$\\
		\cline{2-3} 
		& $^{210}$Bi & 1.6 $\times$ 10$^{-8}$\\
		\cline{2-3} 
		& $^{210}$Po & 1.6 $\times$ 10$^{-6}$\\
		\hline 
		& $^{232}$Th & 5.0 $\times$ 10$^{-4}$\\
		\cline{2-3} 
		& $^{228}$Ra to $^{228}$Ac & 6.98 $\times$ 10$^{-5}$ \\
		\cline{2-3} 
$^{232}$Th  	& $^{228}$Th & 8.3 $\times$ 10$^{-7}$\\
		\cline{2-3} 
		& $^{224}$Ra to $^{208}$Tl & 1.14 $\times$ 10$^{-4}$\\
		\hline\hline
	\end{tabular}
	
  \label{tab:my_label0}
\end{table}

The simulation accounts for the distribution of these
identified contaminants, integrating available experimental
knowledge regarding their presence in the detector
components, shielding materials, and surrounding environment.
In this model, $^{238}$U, $^{232}$Th and $^{235}$U
are confined to the front-end electronics (pre-amplifier) of
the HPGe detector; $^{135}$Xe, $^{60}$Co and $^{54}$Mn are
confined to the air gap between the Cu cells/bricks, and
the NaI(Tl) detector; while $^{137}$Cs is confined to the
CsI(Tl) detector and the ACV detector. A total of 5~million
events were simulated for each detector configuration to
ensure statistically robust estimates of background
contributions. The simulation outputs are stored in
ROOT-compatible files and analyzed using the ROOT data
analysis framework, allowing detailed energy calibration,
spectrum reconstruction, and comparison with experimental
measurements. A threshold cut of 20~keV was applied to
the ACV detector in the simulated dataset to selectively
identify $\gamma$-induced events. Since cosmic-ray induced
events were not included in the GEANT4 framework, the CRV
contribution was excluded from the simulation analysis.

\subsection{Fission activity in the decay chains of $^{238}$U
  and $^{232}$Th}
\label{opt::act}
Several long- and short-lived radionuclides in the $^{238}$U and
$^{232}$Th decay series undergo alpha decay, emitting particles
with energies ranging from 4 to 6~MeV. Other radionuclides in
these series decay by emitting beta particles accompanied by
$\gamma$ rays. Each chain also includes a radioactive isotope of
radon $-$ namely actinon ($^{219}$Rn), thoron ($^{220}$Rn), and
radon ($^{222}$Rn) and ultimately terminates with a stable
isotope of lead.

In practice, the $^{238}$U and $^{232}$Th decay series rarely
maintain complete secular equilibrium owing to variations in
material processing and environmental conditions. It follows
that these chains are divided into sub-chains, each anchored
by radionuclides possessing comparatively long half-lives
that dominate the activity of the respective segment. Furthermore,
to achieve a more accurate understanding of radioactive background
origins, the $^{238}$U decay sequence is partitioned into nine
sub-chains, while the $^{232}$Th series is divided into four
sub-chains. The grouping of these sub-chains, together with their
respective isotopes, is summarized in table~\ref{tab:my_label0}.

To construct a reliable background model and quantify the
contribution of each radioactive component, we follow a
systematic activity-determination procedure in which each
isotope chain, comprising multiple nuclides arising from
successive radioactive decays, is treated as a single
entity for optimization. In our approach, we begin by
assigning a trial activity to the isotope or sub-chain under
consideration. Specifically, for each decay chain,
the highest-energy and most prominent $\gamma$-ray peak
is selected as a reference feature, ensuring a distinct
signature with negligible spectral overlap. The
simulated spectrum is then generated, and the activity
is iteratively adjusted (or optimized) until the
intensity of this reference peak in the simulation
closely reproduces that observed in the experimental
spectrum. Importantly, the optimization is not limited to
reproducing the reference peak alone: the resulting activity
is further constrained to ensure that the overall simulated
spectrum, including both the continuum and other weaker lines,
remains in good agreement with the measured spectrum.
Consequently, an exact one-to-one agreement between the
simulated and measured intensities of the reference peak
is not obligatory. The activity value at which this
agreement is achieved is taken
as the optimized activity. The same optimization step is
carried out for all isotopes and sub-chains, ensuring
that each element of the background model is anchored
to the data via consistency between corresponding peaks
in the simulated and measured spectra. The activity
values mentioned in the upcoming sections originate from
this iterative optimization method.

In extending the optimization procedure to a quantitative
chain-resolved analysis, the decay sub-chains were
treated according to their spectral prominence. Among the
thirteen decay sub-chains considered in the analysis,
only five were independently optimized. These include three
sub-chains from the $^{238}$U series ($^{234}$Th to
$^{234}$Pa, $^{226}$Ra, and $^{222}$Rn to $^{210}$Tl) and
two from the $^{232}$Th series ($^{228}$Ra to $^{228}$Ac
and $^{224}$Ra to $^{208}$Tl). The activity concentrations
of these five sub-chains are optimized independently to
closely reproduce the intensity of the highest-energy
peak of each decay sub-chain with its experimental
counterpart, while simultaneously ensuring good overall
agreement with the measured spectrum. The resulting
optimized activity concentrations are listed in
table~\ref{tab:my_label0}. The remaining eight sub-chains
do not exhibit identifiable peaks in the measured spectrum,
therefore, their activity concentrations are not optimized
independently but are fixed at normalization values to
ensure proper continuum reproduction and overall agreement
with the measured spectrum.

\subsubsection{Confinement of $^{238}$U and $^{232}$Th decay chains
  and $^{235}$U isotope}
The decay chains of $^{238}$U and $^{232}$Th were simulated by
generating events for each of their individual decay chains and
isotopes corresponding to the observed data in the front-end
electronics of the HPGe detector. The activities considered for
each decay chain and isotope are listed in table~\ref{tab:my_label0}.
Additionally, $^{235}$U, another naturally occurring radioactive
nuclide, was also simulated in the front-end electronics of the
HPGe detector. The activity assumed for the isotopes is
7.39$\times10^{-7}$~Bq/kg. 

\begin{figure}[h!]
  \centering
  \includegraphics[width=\textwidth]{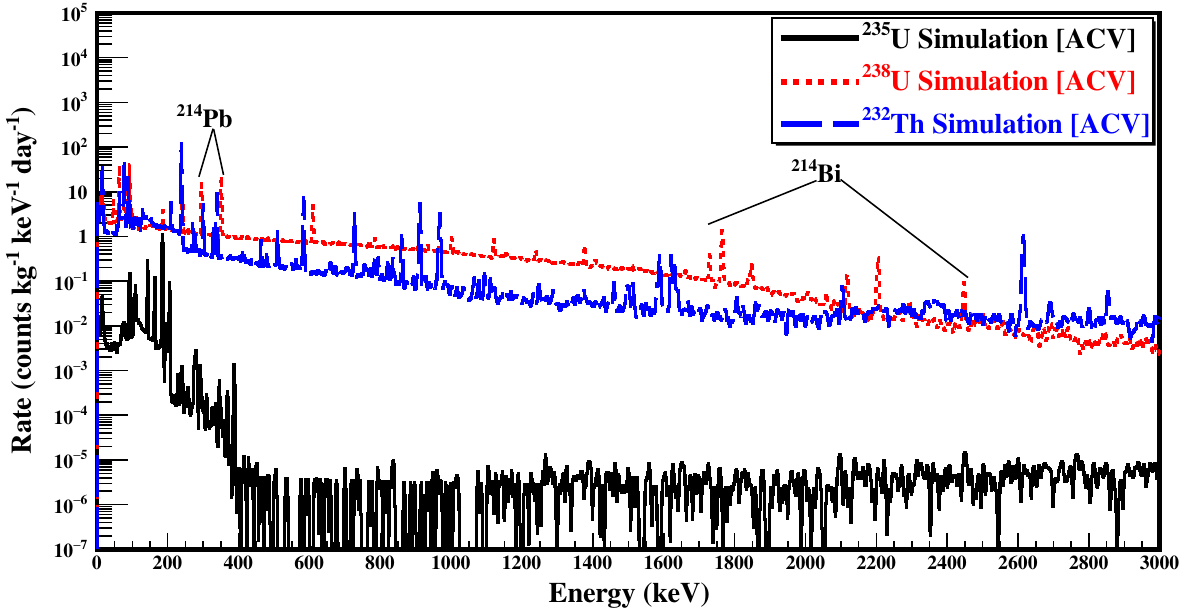} \\
  {\bf (a)} \\
  \includegraphics[width=\textwidth]{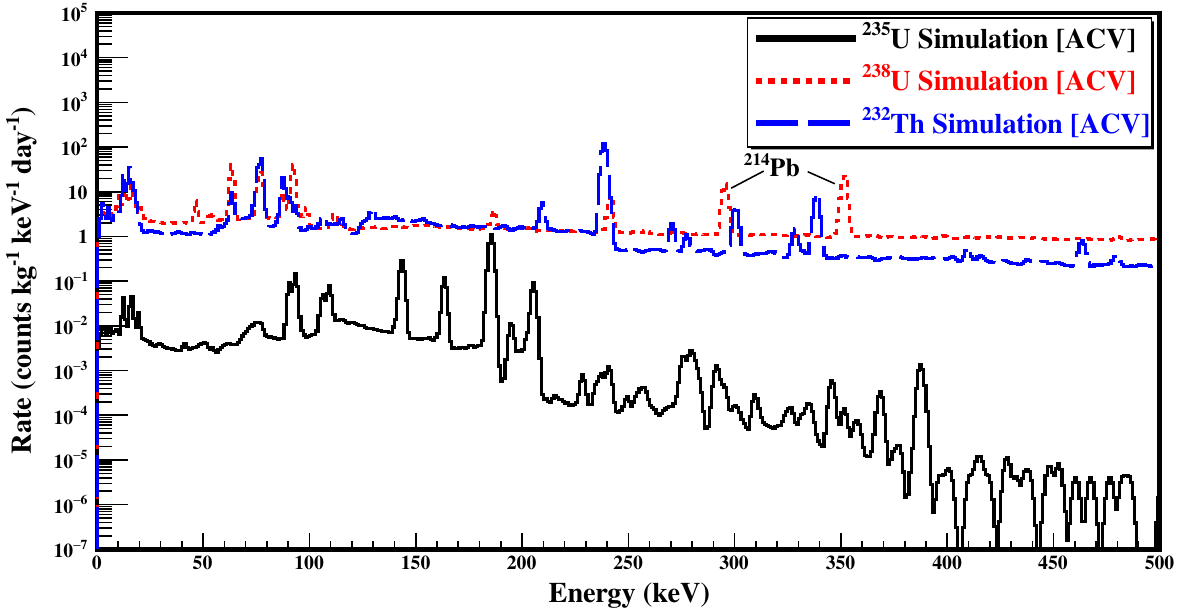} \\
  {\bf (b)}
  \caption{Simulated $\gamma$-ray spectra of
    $^{238}$U, $^{232}$Th, and $^{235}$U isotopes with
    ACV suppression: (a) full spectrum up to 3~MeV,
    and (b) low-energy region below 500~keV.} 
  \label{fig_mom12_STP}
\end{figure}

The simulated decay spectra of $^{238}$U,
$^{232}$Th and $^{235}$U are shown in figure~\ref{fig_mom12_STP}. 
It can be observed that the intensities of
peaks of $^{235}$U are low compared to $^{238}$U
and $^{232}$Th but the presence of multiple discrete peaks
(e.g. 143.8 and 185.7~keV) turns it into a non-negligible
contributor of background in the region below $\sim$ 500~keV).
$^{238}$U and $^{232}$Th show discrete photo peaks reflecting
cascade $\gamma$ emissions and compton-scattered contributions
from their respective decay chains. The $^{238}$U chain
displays a relatively flat continuum across the entire energy
range, with several prominent $\gamma$-ray peaks (e.g., around
609~keV, 1764~keV, and 2204~keV). Its dominance in the
mid-energy region ($500 \lesssim E \lesssim 2000~\text{keV}$)
makes it the principal contributor to long-term background
in low-background detectors. The $^{232}$Th decay series
exhibits a spectral structure comparable to that of the
uranium chain, characterized by distinct $\gamma$-ray features
originating from its daughter isotopes. Among these, the
transition from $^{208}$Tl produces a prominent emission at
2614.5~keV, one of the most intense high-energy $\gamma$-lines
found in natural radioactivity. This line is of particular
concern in rare-event detection experiments, as the
2614.5~keV photon can generate background events that closely
resemble signals of interest, such as the 0$\nu\beta\beta$
of $^{76}$Ge at 2039~keV, thereby complicating spectral
interpretation and necessitating precise background modeling.
The $^{235}$U component, though approximately two orders of
magnitude weaker than the $^{238}$U and $^{232}$Th series,
exhibits a rich structure composed of numerous low-intensity
$\gamma$ transitions, predominantly concentrated in the
low-energy region (0-500~keV) as illustrated in the inset.
Despite its comparatively minor overall contribution, these
closely spaced spectral features can complicate background
discrimination and spectral fitting in the sub-MeV range,
which is particularly critical for reactor neutrino
measurements and low-threshold DM detection experiments.

\begin{figure}[h!]
  \centering 
  \includegraphics[width=\textwidth]{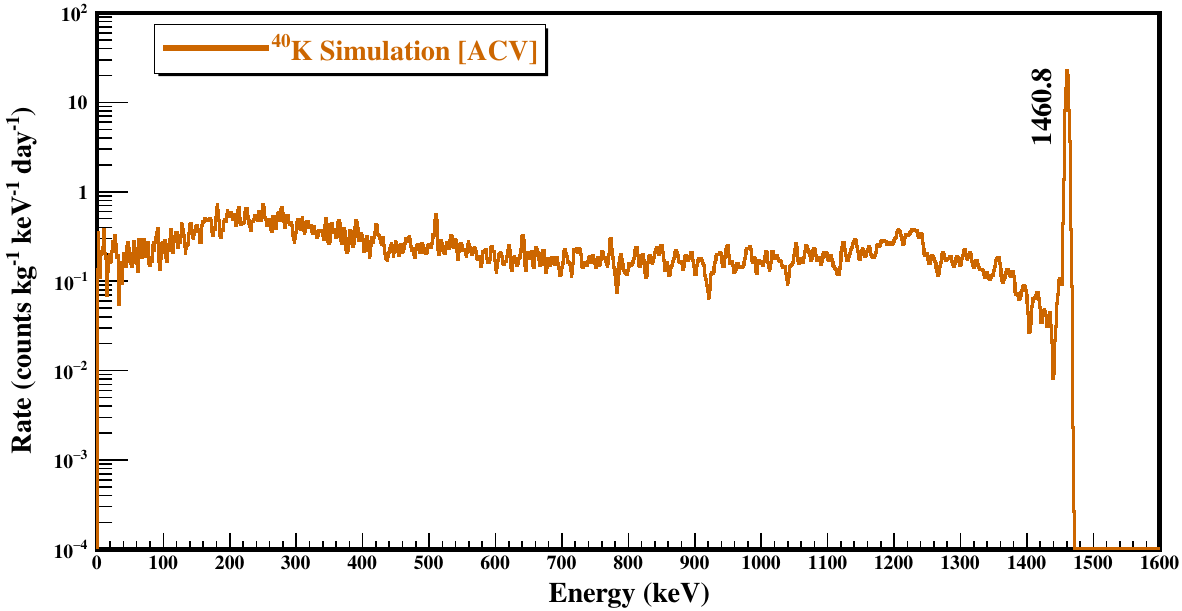}	
  \caption{The simulated $\gamma$-ray spectrum of
    $^{40}$K with ACV selection applied,
    highlighting the impact of anti-Compton
    veto on the detector response. The peak at
    1460.8~keV represents the most prominent
    $\gamma$-line of $^{40}$K, with a relative
    intensity of 10.66\%.} 
  \label{fig_mom6}
\end{figure}

\subsection{Confinement of $^{40}$K isotope}
The isotope $^{40}$K is among the primary sources
of background in the HPGe detector and plays a
significant role in low-background analyses.
It is present in very low concentrations
(less than 0.5~parts per million (ppm) by weight)
in NaI(Tl). Consequently, $^{40}$K is confined to
the NaI(Tl) ACV detector of the setup. The
activity considered in this case is
3.54$\times10^{-7}$~Bq/kg.

The ACV selection suppression spectrum of
$^{40}$K is shown in figure~\ref{fig_mom6}.
It is evident that the spectrum maintains
a nearly uniform background of
$\sim$0.1~counts~kg$^{-1}$keV$^{-1}$day$^{-1}$,
with minor deviations corresponding to absorption
edges or detector-response effects.
A dominant spectral feature emerges at
1460.8~keV, corresponding to the well-known
$\gamma$-line of $^{40}$K. The prominence of
this peak above the relatively featureless
continuum confirms the presence of intrinsic
potassium contamination, a ubiquitous background
in rare-event experiments. The spectrum
emphasizes the dual challenge in low-background
searches: controlling the $^{40}$K peak
while also mitigating the diffuse continuum,
both of which can interfere with the detection
of rare signals such as DM or reactor neutrinos.

\subsection{Confinement of $^{137}$Cs isotope}
\begin{figure}[h!]
  \centering 
  \includegraphics[width=\textwidth]{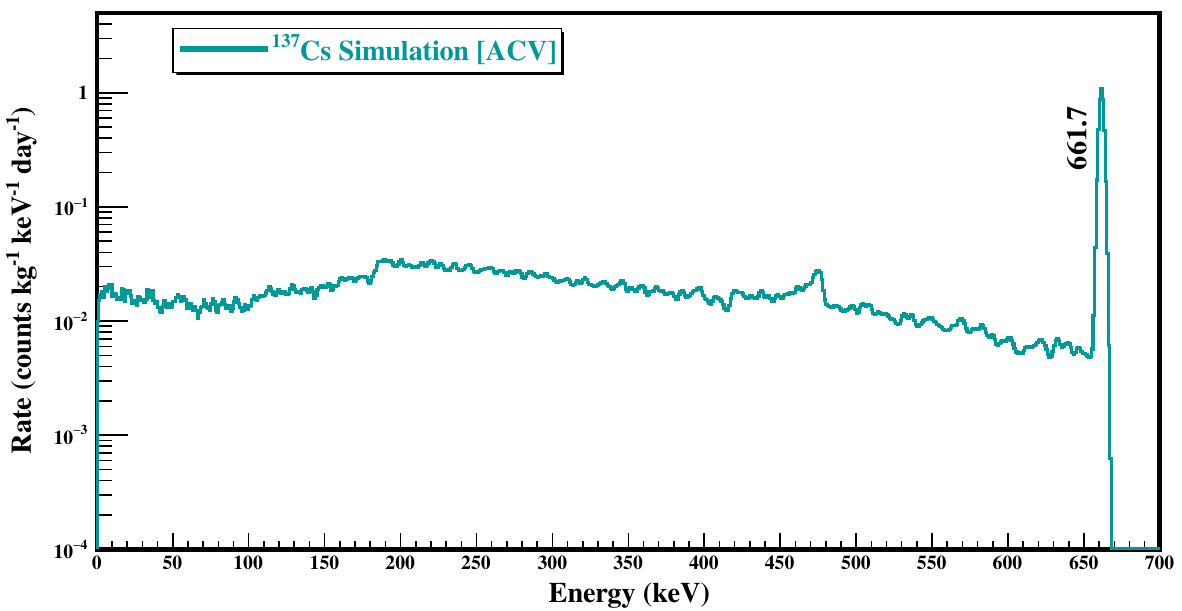}	
  \caption{ACV selection cut simulated $\gamma$-ray
    spectrum of $^{137}$Cs, illustrating the effect of ACV
    suppression on detector response. The peak at
    661.7~keV represents the most prominent $\gamma$-line
    of $^{137}$Cs, with a relative intensity of 85.1\%.} 
  \label{fig_mom7}
\end{figure}

The $^{137}$Cs isotope exists at very low concentration
level in the CsI(Tl) ACV detector and is therefore
restricted to this detector in the background modeling.
The activity considered in this case is
8.55$\times10^{-8}$~Bq/kg. The ACV suppressed spectrum of
$^{137}$Cs is shown in figure~\ref{fig_mom7}.

The presented spectrum in figure~\ref{fig_mom7}
represents a simulated $\gamma$-ray energy
distribution, most likely from a $^{137}$Cs source.
It features a distinct and sharp photo peak at
661.7~keV, corresponding to the characteristic
$\gamma$ emission from the decay of $^{137m}$Ba,
superimposed on a broad Compton continuum
extending toward lower energies. This continuum
arises from Compton scattering processes within
the detector, with a notable inflection near
477~keV marking the expected Compton edge. The
logarithmic intensity scale, spanning several
orders of magnitude, reflects the wide dynamic
range of event rates captured in the simulation.
The absence of additional $\gamma$-lines
indicates a single-isotope source. Overall,
the spectrum reflects the simulated detector
response with the manually applied energy
resolution (following Eq.~\ref{reso}) and
accurate modeling of $\gamma$ interactions,
producing a spectrum consistent with the
expected behavior of a $^{137}$Cs radiation
source.

\subsection{Confinement of isotopes in detector environment}
Among the principal airborne radioactive contaminants
are $^{60}$Co, $^{54}$Mn, and $^{135}$Xe, which are mainly
distributed within the reactor and the detector
environment. Of these, $^{60}$Co and $^{54}$Mn are
significant contributors to the background of the HPGe
detector. These three isotopes are confined within the
air gaps between the Cu end-cap (Cu shell) and the NaI(Tl)
$-$ an ACV detector. The activities considered for
$^{60}$Co, $^{54}$Mn, and $^{135}$Xe are
1.74$\times10^{-3}$~Bq/kg, 4.22$\times10^{-5}$~Bq/kg and
9.67$\times10^{-5}$~Bq/kg, respectively. 

\begin{figure}[h!]
  \centering 
  \includegraphics[width=\textwidth]{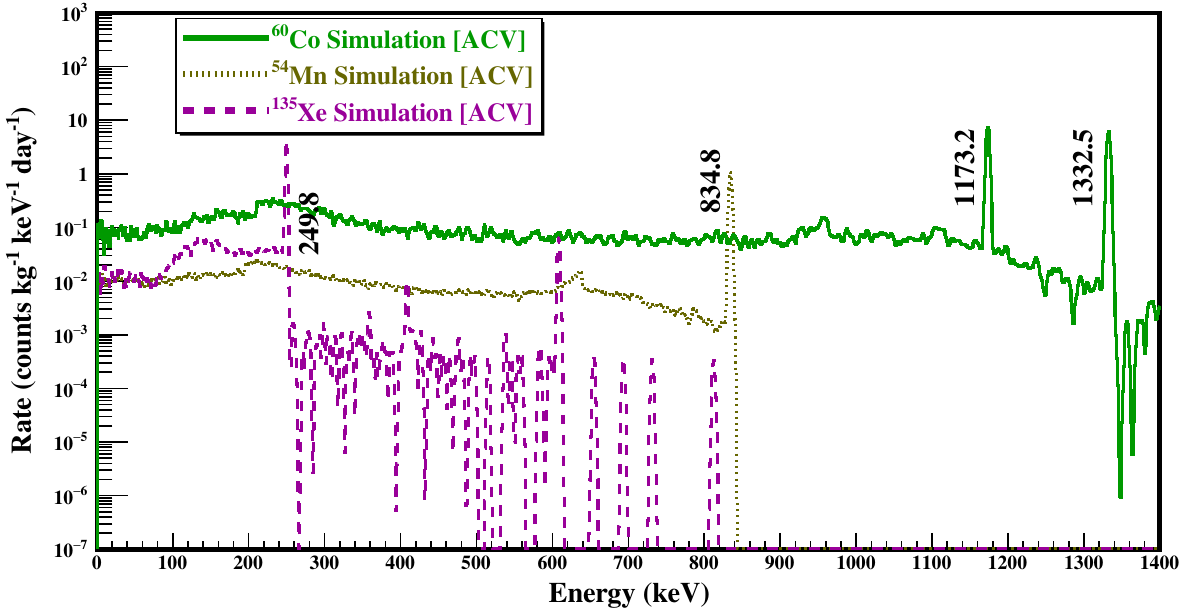}	
  \caption{The MC simulated $\gamma$-ray energy spectra of
    key background contributors, namely $^{135}$Xe (magenta),
    $^{54}$Mn (brown), and $^{60}$Co (green), are shown under ACV
    conditions. The spectra clearly exhibit their
    characteristic full-energy peaks at 249.8~keV ($^{135}$Xe),
    834.8~keV ($^{54}$Mn), and 1173.2~keV and
    1332.5~keV ($^{60}$Co), which are labeled for
    clarity. These spectral features provide clear
    signatures for quantifying isotope-specific background
    identification.}
  \label{fig_mom13_STP}
\end{figure}

The simulated $\gamma$-ray energy spectra for
$^{54}$Mn, $^{135}$Xe, and $^{60}$Co are presented
in figure~\ref{fig_mom13_STP}, highlighting
their characteristic peaks and relative
intensities. The $^{60}$Co spectrum (green curve)
exhibits two prominent photo-peaks at 1173.2~keV
and 1332.5~keV, corresponding to the well-known
cascade $\gamma$ transitions following its
$\beta$-decay to excited states of $^{60}$Ni.
These peaks are accompanied by a broad Compton
continuum extending across lower energies,
typical of multiple scattering events within the
detector medium. The $^{54}$Mn spectrum (brown
curve) displays a dominant single photo-peak at
834.8~keV, characteristic of its decay via
electron capture to an excited state of
$^{54}$Cr, which de-excites by emitting this
monoenergetic $\gamma$ photon. In contrast, the
$^{135}$Xe spectrum (magenta curve) shows a
comparatively weak intensity distribution with
a small but distinct peak at 249.8~keV,
representing its principal $\gamma$ emission.
The logarithmic intensity scale reveals the
wide dynamic range among the isotopes,
emphasizing the much higher photon yield and
interaction probability of $^{60}$Co compared
to $^{54}$Mn and $^{135}$Xe. Overall, the
simulation effectively captures the spectral
features of each isotope, with the relative
activities manually introduced to normalize
the spectra, thereby enabling accurate
modeling of photon-matter interactions and the
detector response across a broad energy range.

\section{Combined spectrum from individual confined isotopic sources}
\begin{figure}[h!]
  \centering 
  \includegraphics[width=\textwidth]{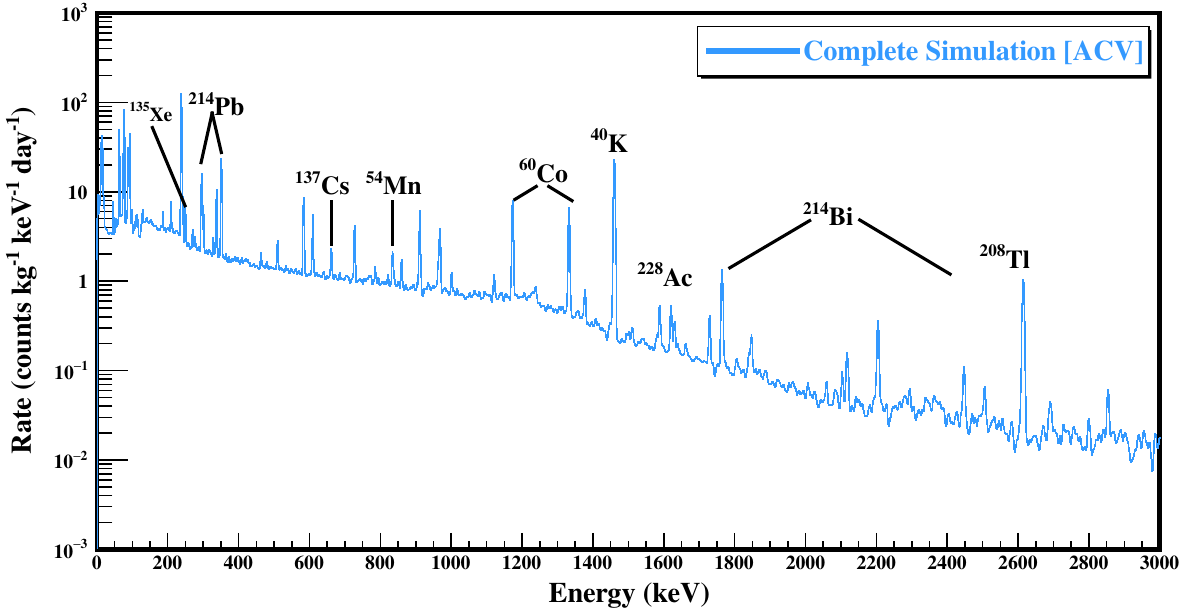}	
  \caption{Combined simulated spectrum of all
    isotopes and decay chains after applying the ACV
    selection cut, with appropriate labeling of the
    relevant peaks.} 
  \label{fig_mom11}
\end{figure}

The complete simulated spectrum is generated by
combining the individual spectra of all confined
isotopes and decay chains, following their
detailed modeling within the detector system.
The final simulated spectrum, including
appropriately labeled peaks corresponding to key
isotopic contributions, is depicted in
figure~\ref{fig_mom11}.

The presented spectrum depicts a simulated environmental
$\gamma$-ray background, comprising multiple naturally
occurring and anthropogenic radionuclides. Prominent peaks
are observed for $^{40}$K, $^{208}$Tl, $^{214}$Bi, $^{228}$Ac, and
$^{60}$Co, each corresponding to their characteristic $\gamma$
transitions. The intense peak near 1460~keV is attributed to
$^{40}$K, a ubiquitous terrestrial isotope contributing
significantly to the natural background. The doublet of
high-energy peaks around 1764~keV and 2614~keV correspond
to $^{214}$Bi and $^{208}$Tl, respectively, key progeny in
the $^{238}$U and $^{232}$Th decay chains, highlighting the
influence of primordial radioisotopes within the simulated
environment. Additional spectral features around 1173~keV
and 1332~keV arise from $^{60}$Co, indicative of
anthropogenic contamination or activation products. The
smaller peaks near 911~keV and 969~keV are associated with
$^{228}$Ac, reflecting contributions from the thorium
series. The logarithmic scale of the intensity axis spans
several orders of magnitude, revealing both dominant and
trace isotopic contributions with high dynamic resolution.
Overall, the spectrum captures the complex interplay of
natural and man-made $\gamma$ emitters, demonstrating a
realistic and well-calibrated simulation of environmental
background radiation, with accurate modeling of $\gamma$-ray
emission probabilities and photon interaction processes. The
evaluated background components associated with the reactor
environment and with the NaI(Tl) and CsI(Tl) ACV detector
assemblies are summarized in table~\ref{tab:my_label1},
providing a quantitative overview of their relative
contributions to the total background.

The contributions of
$\gamma$-ray peaks from $^{238}$U, $^{232}$Th, and $^{235}$U
isotopes are listed and summarized in
table~\ref{tab:my_label2}. Although most isotopes
exhibit a single prominent $\gamma$-ray emission
line, the simulated spectra for the $^{238}$U and
$^{232}$Th decay series are treated as 13 distinct
sub-chains, with secular equilibrium assumed
within each sub-chain, as commonly adopted in
radiation background
modeling~\citep{SeclE::HG::1995,SeclE::GFKnl::2010}.
This assumption implies that
the activity of each daughter nuclide is equal to
that of its parent. However, in practical
experimental conditions, equilibrium may be partially
broken due to radon emanation, material handling,
or surface contamination. Consequently, the
simulated intensities of certain $\gamma$-ray
peaks may differ slightly from the measured values.
Combined with the fact (as also illustrated in
Section~\ref{opt::act}) that only five sub-chains
are independently optimized by matching the
highest-energy $\gamma$-ray peak in the simulation
to the corresponding peak in the experimental data.
The remaining eight sub-chains are fixed at
normalization values to reproduce the continuum.
Since the optimization is intended to achieve
overall agreement between the simulated and measured
spectra rather than exact peak-to-peak matching,
these factors collectively account for the small
differences observed between simulation and
measurement.

This optimization criterion can be quantitatively
demonstrated through representative decay sub-chain
comparisons within the measured and simulated spectra.
For instance, within the $^{238}$U decay series,
the sub-chain $^{222}$Rn to $^{210}$Tl produces
highest-energy $\gamma$-line at 2447.9~keV from its
decay product $^{214}$Bi, demonstrates close agreement
between the measured (0.5$\pm$0.1)
counts~kg$^{-1}\cdot$day$^{-1}$ and the simulated
intensity of 0.6 counts~kg$^{-1}\cdot$day$^{-1}$. In
contrast, the simulated peak intensities associated
with the $^{234}$Th to $^{234}$Pa (1001.0~keV) and
$^{226}$Ra (186.2~keV) sub-chains do not reproduce the
measured values with the same level of proximity. A
similar situation is observed for sub-chains within
the $^{232}$Th series, where the optimization is
governed by global spectral agreement rather than
strict one-to-one peak matching. In this case, the
decay sub-chain $^{228}$Ra to $^{228}$Ac produces
the highest-energy $\gamma$-line at 1630.6~keV, which
exhibits close agreement between the measured
intensity (0.6$\pm$0.1) counts~kg$^{-1}\cdot$day$^{-1}$
and the simulated value of 0.7
counts~kg$^{-1}\cdot$day$^{-1}$. In contrast, the
sub-chain $^{224}$Ra to $^{208}$Tl, with its 2614.5~keV
$\gamma$-line, does not reproduce the measured
intensity with the same level of agreement.

\setlength{\tabcolsep}{0.9em} 
\begin{table}[h!]
  \centering
  \caption{Identified $\gamma$ energies of different
    radioactive isotopes in the simulated spectrum,
    originating from contributions of the reactor
    environment and the NaI(Tl) and CsI(Tl) ACV
    detectors. To enable a clear measurement-simulation
    comparison, the intensities of the measured
    peaks~\citep{Wong::2007} and their simulated
    counterparts are listed jointly in the same column.}
    \renewcommand{\arraystretch}{1.2}
    \begin{tabular}{c|c|c|c|c|c}
    \hline\hline
          Energy  &           & Source/Decay & \multicolumn{2}{c|}{Intensity (kg$^{-1}$day$^{-1}$)} & \\ \cline{4-5}
          (keV)   & Isotopes  & Series       & Meas.          & Sim.                                  & Location          \\ \hline
          249.8   & $^{135}$Xe & Environment  & 11.6$\pm$0.5   & 11.7                                  & Air gap           \\ 
          661.7   & $^{137}$Cs & CsI(Tl)      & 4.6$\pm$0.2    & 4.7                                  & CsI(Tl) ACV       \\ 
          834.8   & $^{54}$Mn  & Environment  & 3.6$\pm$0.3    & 3.6                                   & Air gap           \\ 
          1173.2  & $^{60}$Co  & Environment  & 26.0$\pm$0.3   & 27.1                                  & Air gap           \\ 
          1332.5  & $^{60}$Co  & Environment  & 27.0$\pm$0.3   & 27.8                                  & Air gap           \\ 
          1460.8  & $^{40}$K   & NaI(Tl)      & 106.4$\pm$1.0  & 108.3                                  & NaI(Tl) ACV       \\ \hline\hline
         
    \end{tabular}
    
    \label{tab:my_label1}
\end{table}

\setlength{\tabcolsep}{0.6em}
\begin{table}[h!]
  \centering
  \caption{The contributions of $\gamma$-ray peaks from 
    $^{238}$U, $^{232}$Th, and $^{235}$U isotopes to the
    complete simulated spectrum are summarized. The intensities
    obtained from the measurement~\citep{Wong::2007} and from
    the simulation are presented together to enable a
    straightforward comparison.}
  \renewcommand{\arraystretch}{1.0}
    \begin{tabular}{c|c|c|c|c|c}
    \hline\hline
    Energy    &            & Source/Decay   & \multicolumn{2}{c|}{Intensity (kg$^{-1}$day$^{-1}$)} & \\ \cline{4-5} 
    (keV)     & Isotopes   & series         & Meas.           & Sim.                             & Location              \\ \hline
    92.6      & $^{234}$Th  & $^{238}$U       & 11.9$\pm$0.5    & 62.2                             &                       \\ 
    143.8     & $^{235}$U   & $^{235}$U       & 5.1$\pm$0.8     & missing                          &                       \\ 
    185.7     & $^{235}$U   & $^{235}$U       & 17.2$\pm$0.4    & 8.9                              &                       \\ 
    186.2     & $^{226}$Ra  & $^{238}$U       & 17.2$\pm$0.4    & 8.9                              &                       \\ 
    238.6     & $^{212}$Pb  & $^{232}$Th      & 18.8$\pm$0.5    & 346.9                            &                       \\ 
    295.2     & $^{214}$Pb  & $^{238}$U       & 6.3$\pm$0.3     & 4.3                              &                       \\ 
    338.3     & $^{228}$Ac  & $^{232}$Th      & 3.7$\pm$0.5     & 2.4                              &                       \\ 
    351.9     & $^{214}$Pb  & $^{238}$U       & 17.1$\pm$0.4    & 58.3                             &                       \\ 
    463.0     & $^{228}$Ac  & $^{232}$Th      & 1.6$\pm$0.3     & 1.7                              &                       \\ 
    583.2     & $^{208}$Tl  & $^{232}$Th      & 14.4$\pm$0.3    & 25.5                             &                       \\ 
    609.3     & $^{214}$Bi  & $^{238}$U       & 8.1$\pm$0.2     & 13.9                             &                       \\ 
    727.3     & $^{212}$Bi  & $^{232}$Th      & 6.4$\pm$0.2     & 11.2                             &                       \\ 
    766.4     & $^{234m}$Pa & $^{238}$U       & 5.0$\pm$0.3     & 1.3                              &                       \\ 
    785.4     & $^{212}$Bi  & $^{232}$Th      & 1.7$\pm$0.4     & 1.7                              &                       \\ 
    786.0     & $^{214}$Pb  & $^{238}$U       & 1.7$\pm$0.4     & 1.7                              &                       \\ 
    795.0     & $^{228}$Ac  & $^{232}$Th      & 2.7$\pm$0.8     & 1.2                              &                       \\
    860.6     & $^{208}$Tl  & $^{232}$Th      & 3.5$\pm$0.3     & 3.5                              &                       \\ 
    911.2     & $^{228}$Ac  & $^{232}$Th      & 19.1$\pm$0.3    & 19.5                             & Front-end electronics \\ 
    964.8     & $^{228}$Ac  & $^{232}$Th      & 14.4$\pm$0.3    & 14.4                             &                       \\ 
    969.0     & $^{228}$Ac  & $^{232}$Th      & 14.4$\pm$0.3    & 14.4                             &                       \\ 
    1001.0    & $^{234m}$Pa & $^{238}$U       & 11.4$\pm$0.3    & 2.0                              &                       \\ 
    1120.3    & $^{214}$Bi  & $^{238}$U       & 6.7$\pm$0.5     & 2.8                              &                       \\ 
    1238.1    & $^{214}$Bi  & $^{238}$U       & 1.2$\pm$0.2     & 1.1                              &                       \\ 
    1377.7    & $^{214}$Bi  & $^{238}$U       & 1.9$\pm$0.3     & 1.8                              &                       \\ 
    1509.2    & $^{214}$Bi  & $^{238}$U       & 0.6$\pm$0.1     & 0.4                              &                       \\ 
    1588.2    & $^{228}$Ac  & $^{232}$Th      & 2.5$\pm$0.1     & 1.7                              &                       \\ 
    1620.5    & $^{212}$Bi  & $^{232}$Th      & 1.6$\pm$0.1     & 1.6                              &                       \\ 
    1630.6    & $^{228}$Ac  & $^{232}$Th      & 0.6$\pm$0.1     & 0.7                              &                       \\ 
    1729.6    & $^{214}$Bi  & $^{238}$U       & 1.1$\pm$0.1     & 1.2                              &                       \\ 
    1764.5    & $^{214}$Bi  & $^{238}$U       & 5.9$\pm$0.9     & 6.0                              &                       \\ 
    1847.4    & $^{214}$Bi  & $^{238}$U       & 0.7$\pm$0.3     & 0.7                              &                       \\ 
    2118.6    & $^{214}$Bi  & $^{238}$U       & 0.2$\pm$0.1     & 0.7                              &                       \\ 
    2204.2    & $^{214}$Bi  & $^{238}$U       & 2.3$\pm$0.4     & 2.2                              &                       \\  
    2447.9    & $^{214}$Bi  & $^{238}$U       & 0.5$\pm$0.1     & 0.5                              &                       \\  
    2614.5    & $^{208}$Tl  & $^{232}$Th      & 14.5$\pm$0.2    & 6.8                              &                       \\ \hline\hline
    \end{tabular}
    \label{tab:my_label2}
\end{table}

\section{Cross-comparison of simulated and measured spectra}
\begin{figure}[h!]
  \centering
  \includegraphics[width=\textwidth]{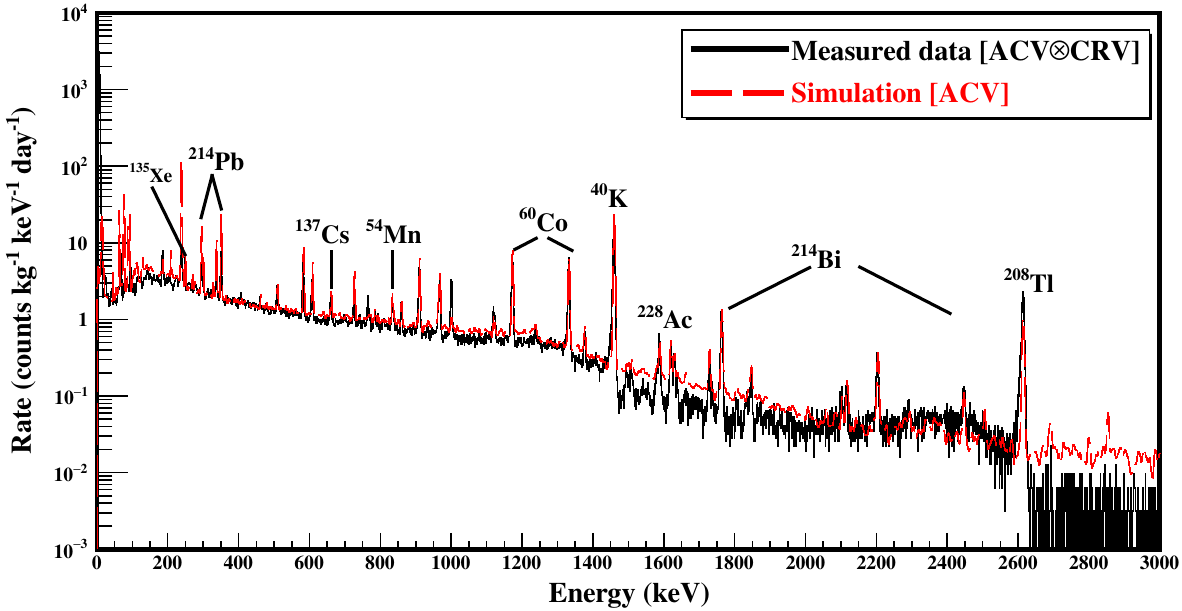} \\
  {\bf (a)} \\
  \includegraphics[width=\textwidth]{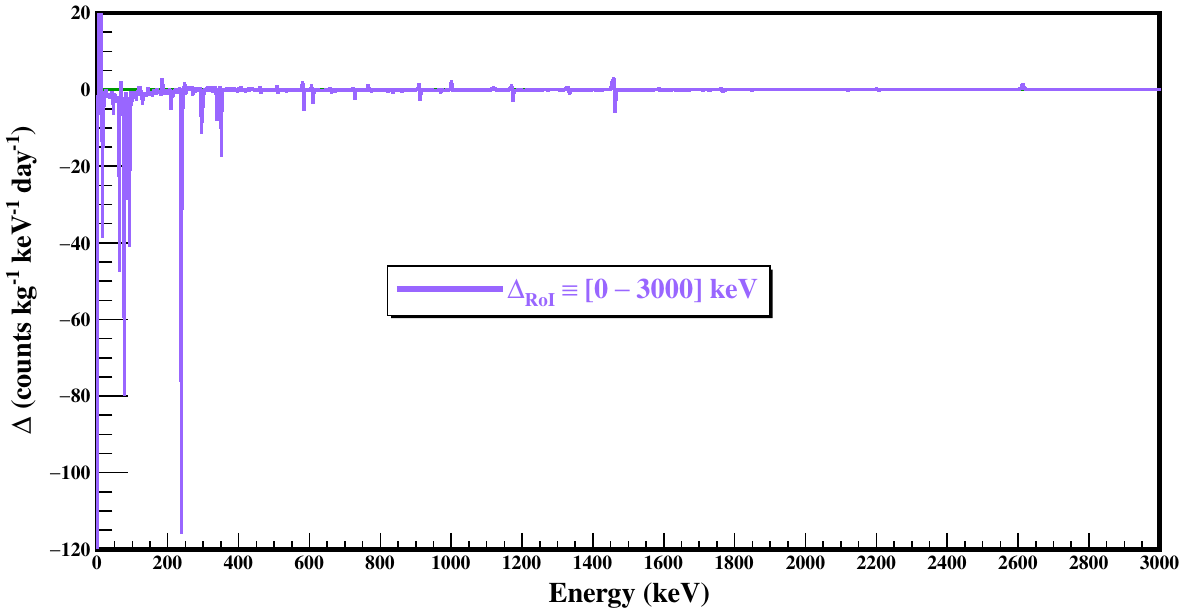} \\
  {\bf (b)}
  \caption{(a) Experimental measurements compared with the total
    simulated spectrum derived from the sum of individual confined
    isotopic source spectra, illustrating the correspondence
    between simulation and observed data across the energy range.
    (b) Complete differential spectrum ($\Delta\equiv$ measured$-$simulated)
    spanning the energy range $\Delta_{\rm{RoI}}$ up to 3~MeV, used to assess the
    fidelity of the simulation relative to experimental data.} 
  \label{fig::Com::Spec}
\end{figure}

The complete simulated spectrum is compared with the
experimentally measured spectrum. The activity of each
decay chain or isotope is optimized so that the
intensity of a peak present in simulated spectrum
will be matched with the intensity of corresponding
peak present in experimental spectrum to get the
similar intensity. The energy resolution of the HPGe
detector is incorporated manually and uniformly for
all simulated spectral lines using the energy
dependent Gaussian smearing, as formulated in
Eq.~\ref{reso}. This consistent application ensures
that the relative heights and widths of all peaks are
maintained and that composite peaks resulting from
multiple transitions are accurately represented.
For each simulated decay spectrum, the ACV selection
cut has been applied with a threshold of 20~keV. Each
individual simulated spectrum has an energy range from
0 to 3~MeV. The combined simulated and experimental
spectra covering the full energy range up to 3~MeV are
shown in figure~\ref{fig::Com::Spec}.

The comparison between the experimental data (black) and
the MC simulation (red) demonstrates a strong overall
agreement, indicating that the simulation reliably
reproduces the detector's physical response and the main
background contributions present in the measured spectrum.
Both spectra exhibit the expected behavior of a decreasing
continuum with increasing energy, superimposed by several
distinct $\gamma$-ray peaks that correspond to
characteristic emissions from naturally occurring
radioisotopes such as $^{40}$K (1460~keV), $^{214}$Bi
(1764~keV), and $^{208}$Tl (2615~keV). The good alignment
of these features in both intensity and energy position
confirms accurate energy calibration, proper modeling of
detector geometry, and well-represented physical processes
such as Compton scattering and photoelectric absorption.

The structures observed in the experimental spectrum near
770~keV and around 1~MeV are attributed to $\gamma$-ray
emissions from the decay of $^{234m}$Pa, which is part of
the $^{238}$U decay chain. These $\gamma$-lines are also
present in the simulated spectra, confirming that their
physical origin is correctly included in the background
model. The apparent difference in visibility arises from
the activity normalization applied in the simulation, which
is tuned to reproduce the overall spectral shape and
continuum level. As a consequence, while the energies of
the $^{234m}$Pa lines are accurately reproduced, their
intensities are slightly underestimated in the simulation.
This does not indicate a missing background component, but
rather reflects the chosen global normalization strategy.

Slight deviations are observed, particularly at lower
energies, where the simulation tends to overestimate the
rate, and at higher energies near the 2614~keV peak, likely
due to differences in veto system implementation (the
CRV not included in simulation), uncertainties in
shielding composition, or limited counting statistics
in the data. In the current simulation, the intensity of
$^{208}$Tl line at the 2614~keV is underestimated with
respect to the experimental spectrum, despite the activity
normalization of the $^{232}$Th decay chain. The activity
normalization was primarily constrained using the
low-energy region (0$-$1 MeV), where the statistical
precision is higher and multiple lines contribute
simultaneously. This region contains multiple
$\gamma$-ray peaks arising from successive decays within
the $^{232}$Th series. In the simulation, the series is
treated as four sub-chains, of which two, $^{228}$Ra to
$^{228}$Ac and $^{224}$Ra to $^{208}$Tl, exhibit
identifiable peak contributions and are therefore
optimized independently using their highest-energy,
well-resolved $\gamma$-lines, whereas the remaining two
sub-chains lack distinct peaks in the measured spectrum
are constrained by normalization to maintain continuum
agreement. While this procedure yields good agreement
at low energies, a residual mismatch remains at high
energies, most notably in the intensity of the 2614~keV
line. This mismatch indicates limitations in the present
background modeling at high energies, which may arise from
a combination of factors, including uncertainties in the
spatial distribution of $^{232}$Th bearing materials,
incomplete modeling of surrounding passive materials, and
residual imperfections in the treatment of high-energy
$\gamma$-ray interactions and detector response.
Despite these small discrepancies, the comparison
demonstrates a broadly consistent correspondence between
data and simulation across several orders of magnitude, validates the
robustness of the simulation framework, confirming its
suitability for detailed background characterization,
detector performance optimization, and predictive
modeling in low-background experiments.

\begin{figure}[h!]
  \centering 
  \includegraphics[width=0.48\textwidth]{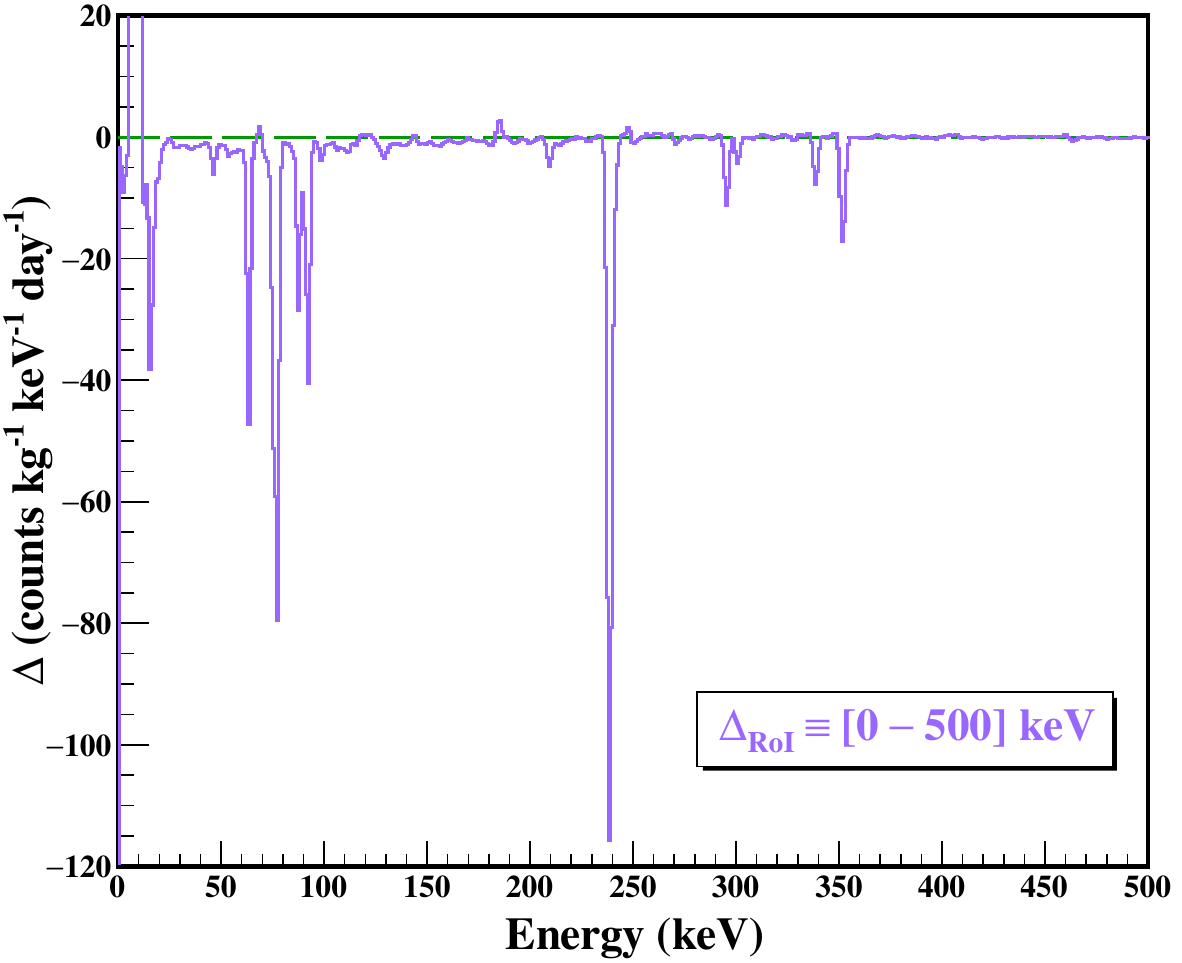}
  \includegraphics[width=0.48\textwidth]{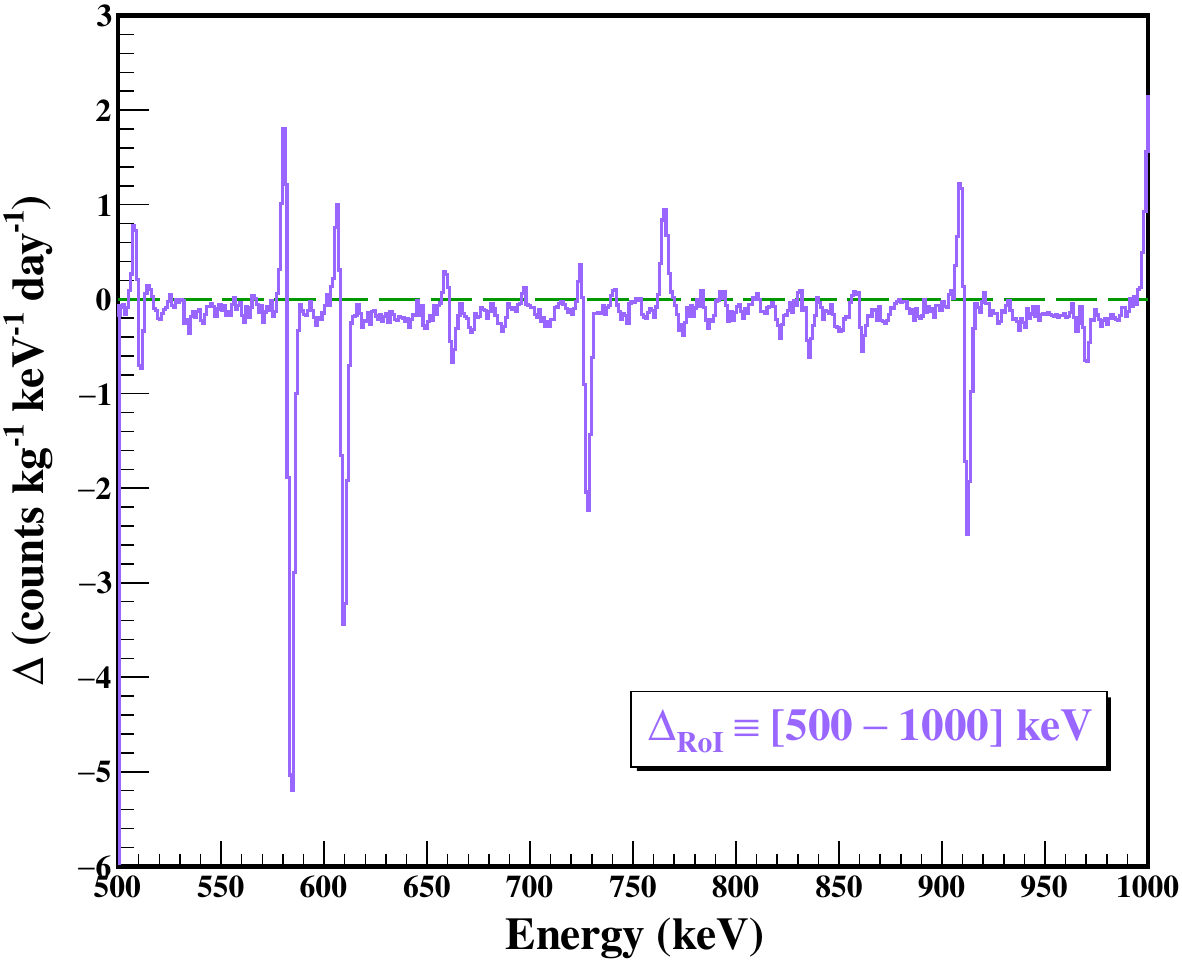} \\
  {\bf (a)}        \hspace{7.5cm}    {\bf (b)} \\
  \includegraphics[width=0.48\textwidth]{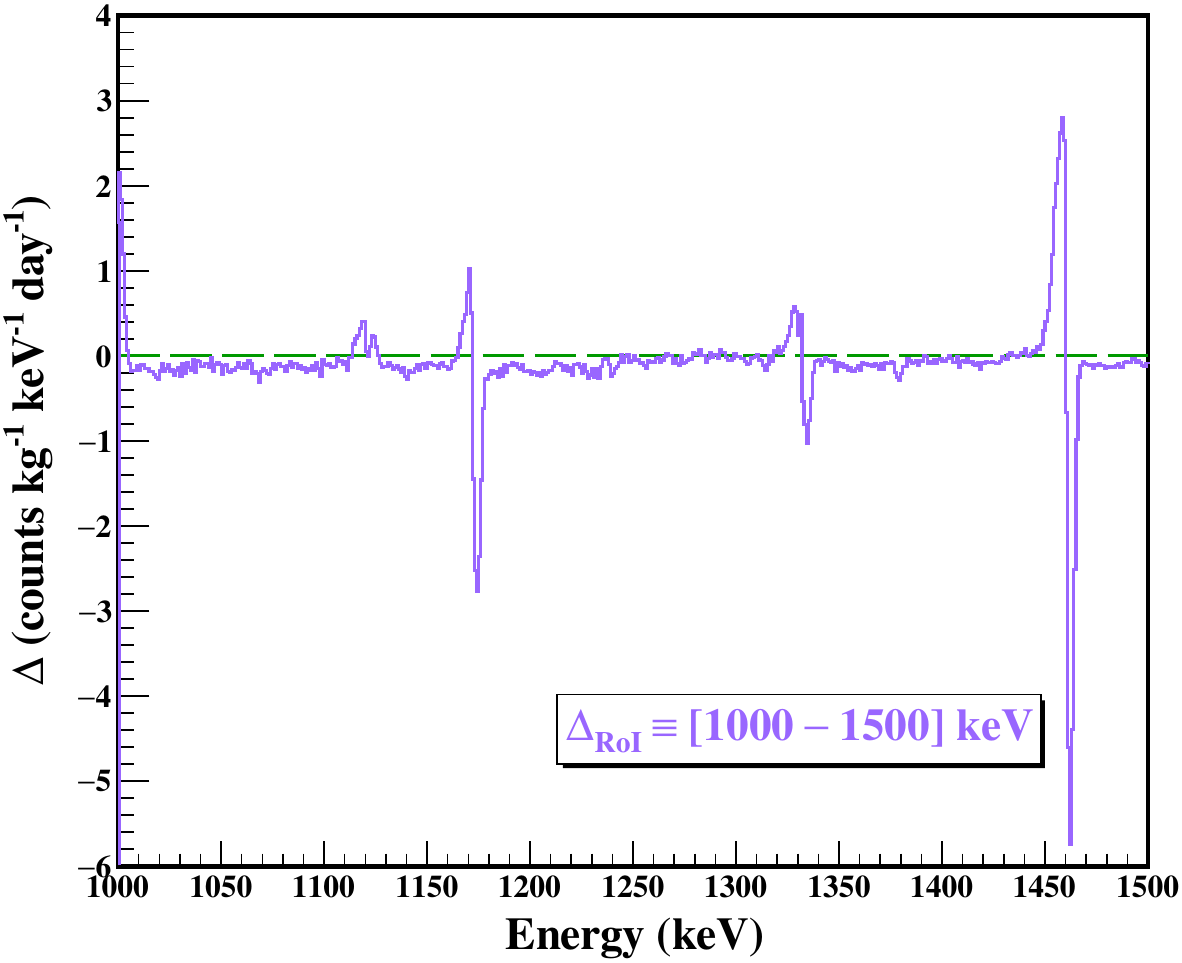}
  \includegraphics[width=0.48\textwidth]{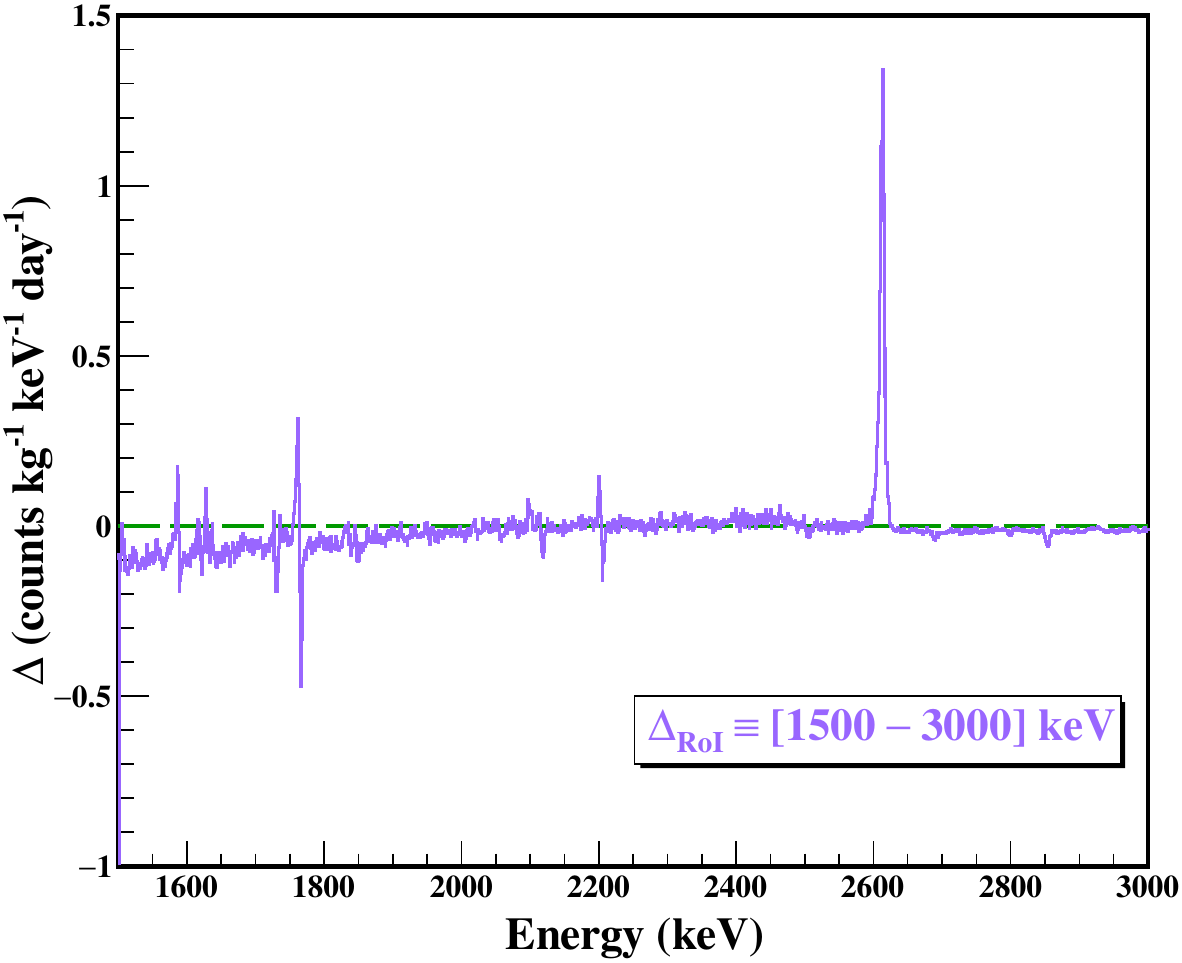} \\
  {\bf (c)}        \hspace{7.5cm}    {\bf (d)} 
  \caption{Differential spectra ($\Delta\equiv$ observation$-$simulation)
    for the selected four energy ranges ($\Delta_{\rm{RoI}}$): (a) (0-500)~keV,
    (b) (500-1000)~keV, (c) (1000-1500)~keV, and (d) (1500-3000)~keV.
    These segmented spectra emphasizes the agreement between
    simulation and measurement and identify regions with minor
    deviations, particularly at known $\gamma$-ray peaks.} 
  \label{fig_mom113}
\end{figure}

The differential spectrum ($\Delta\equiv$ observed$-$simulated) is
depicted in figure~\ref{fig::Com::Spec}(b), allowing
assessment of agreement between measurement and simulation.
The presented difference spectrum illustrates the variation
in $\gamma$-ray count rates across the energy range of
(0-3000)~keV, providing a comparative assessment between
simulated and measured spectral datasets. The straight line,
indicated by the green dashed line, serves as a reference and
demonstrates that, over most of the energy range, the rate
difference remains negligible, with no significant deviation,
reflecting good agreement and consistency between the
compared spectra. However, noticeable negative excursions below 500~keV suggest a
significant reduction in detected counts within the
low-energy region, which may arise from enhanced absorption,
detector threshold effects, or differences in background
subtraction methodologies. The pronounced dips in this
region could also reflect systematic discrepancies associated
with low-energy photon attenuation or electronic noise
filtering during the data acquisition process. Beyond
approximately 600~keV, the rate difference stabilizes at
vanishingly small values ($\mathcal{O}$(0.1)~counts~kg$^{-1}$keV$^{-1}$day$^{-1}$),
implying uniform detector response and effective
spectral alignment at higher energies. Overall, the
difference spectrum indicates good agreement between the two
datasets over a large fraction of the analyzed energy range,
while revealing noticeable deviations at low energies. These
deviations point to limitations in the current material
modeling and the assumed spatial distribution of certain
radioactive isotopes. The simulation framework provides
direct access to the origin of these features, enabling a
systematic investigation and refinement of the detector
geometry and background model in the low-energy region.

To facilitate clear visualization of the differences
in peak rates, the complete difference spectrum is
divided into four energy intervals ($\Delta_{\rm{RoI}}$): (0-500)~keV,
(500-1000)~keV, (1000-1500)~keV, and (1500-3000)~keV.
The corresponding segmented spectra are displayed in
figure~\ref{fig_mom113}. This segmentation allows small
but physically relevant deviations, particularly in the
higher-energy regions, to be more clearly resolved.

In the low-energy region
(0-100~keV), as shown in figure~\ref{fig_mom113}(a), the
comparison between the measured and simulated spectra reveals
three additional peaks present exclusively in the simulation,
located at 63.3~keV, 77~keV, and 87~keV. These lines originate
from $^{234}$Th (3.7\% intensity) and the x-ray
transitions of $^{214}$Pb (8.7\%) and $^{212}$Pb (3.91\%),
respectively. The 63.3~keV line is identified as the
$\gamma$-ray from the decay of $^{234}$Th to $^{234}$Pa,
while the 77~keV and 87~keV lines originate from
characteristic atomic x-rays of $^{214}$Pb and $^{212}$Pb in
the $^{238}$U and $^{232}$Th decay chains, respectively,
following $\beta$-decay and internal conversion. Their
absence in the measured spectrum may result from detector
threshold effects, finite energy resolution, passive shielding
near SCGe crystal or veto suppression. Further
investigation is underway to ascertain whether these
features are experimentally suppressed or physically
absent.

In the high-energy region, as shown in
figures~\ref{fig_mom113}(b)$-$(d), the apparent structures near
intense peaks arise from residual differences in peak shape,
primarily due to finite statistics and minor mismatches in the
modeling of the detector energy resolution, rather than from a
global miscalibration. We emphasize that the energy calibration
is consistently applied to both data and simulation, and that
the remaining deviations in the difference spectra are localized
and of limited magnitude. Excluding these small deviations, the
differential spectra in the high-energy region remain close to
minimal values, indicating consistent agreement between the
simulated and measured spectra.

\section{Conclusions}
A comprehensive understanding of background sources is
essential for all experiments and becomes particularly
critical in rare-event searches, where background
reduction directly determines the achievable sensitivity.
In reactor-based neutrino experiments, reactor-off data
are typically used to evaluate background contributions. 
Nonetheless, in commercial reactor facilities like KSNL,
reactor-off periods are considerably shorter than
reactor-on operations, resulting in limited statistical
precision. Consequently, accurate and comprehensive
background modeling becomes essential for reliable
background characterization. The recent background modeling
approach adopted by the CONUS+
experiment~\citep{Ackerman::5206}, which played a pivotal
role in the first observation of reactor neutrino-induced
CE$\nu$NS, exemplifies the critical importance of precise
background characterization. 

In the present work, a detailed GEANT4-based simulation
is developed to replicate the TEXONO detector setup at
KSNL, including its geometry, materials, and shielding
structures. This framework enables realistic estimation
of both intrinsic and environmental background
contributions. All significant radioactive isotopes
observed in the experimental spectra are incorporated
into the model at physically motivated locations, allowing
accurate reproduction of the measured background and
enhanced understanding of its origin. The present simulation
framework developed for the TEXONO experiment provides a
systematic approach to understanding and modeling
background contributions, thereby reducing dependence on
reactor-off data and complementing experimental studies
across the full energy range from keV~to~MeV. The principal
results and scientific implications of this work are
summarized below to highlight the essential conclusions
drawn from the study:

\vspace*{-2 mm}
\begin{enumerate}
\item Naturally occurring radionuclides such as $^{238}$U,
  $^{232}$Th, and $^{235}$U were introduced into the front-end
  electronics (pre-amplifier) of the HPGe detector, a region
  known to harbor trace impurities from manufacturing
  materials. The $^{235}$U isotope contributes a continuum
  feature between 20~keV and 60~keV, adding nearly
  4$\times10^{-3}$~counts~kg$^{-1}$keV$^{-1}$day$^{-1}$ to the
  background. Conversely, $^{238}$U maintains a comparatively
  smoother spectrum with sustained emission rates at the
  $\mathcal{O}$(0.01-1)~counts~kg$^{-1}$keV$^{-1}$day$^{-1}$
  throughout the (200-2500)~keV region, driven by the dense
  $\gamma$-cascade of its long-lived progeny, notably
  $^{214}$Bi and $^{214}$Pb. The $^{232}$Th follows a similar
  continuum but exhibits decreased spectral activity above
  250~keV and contributes
  (8.5$\times10^{-3}$-0.5)~counts~kg$^{-1}$keV$^{-1}$day$^{-1}$
  in the energy region (250-2000)~keV. The analysis confirms
  that the dominant background contributions originate from
  the decay chains of $^{238}$U and $^{232}$Th present in the
  front-end electronics of the HPGe detector.
  
\vspace*{-2 mm}
\item The NaI(Tl) ACV detector was simulated with trace
  quantities of $^{40}$K, estimated to be below 0.5~ppm
  by weight, consistent with the natural abundance in
  NaI crystals. The single line present at 1460.8~keV
  above the continuum confirms mono-energetic emission. The
  Compton continuum-extending across the sub-MeV range
  with a nearly uniform rate of
  0.1~counts~kg$^{-1}$keV$^{-1}$day$^{-1}$ results from
  partial energy deposition from $\gamma$-photons within
  the detector volume. In case of  $^{137}$Cs isotope, the
  Compton continuum spreads with an average rate
  $\approx$ 2$\times10^{-2}$~counts~kg$^{-1}$keV$^{-1}$day$^{-1}$
  below 400~keV region. It follows that minor yet
  observable background contributions are traced to
  $^{40}$K in the NaI(Tl) ACV detector and $^{137}$Cs in the
  CsI(Tl) scintillator. Particularly, in the residual
  spectrum, $^{40}$K $\gamma$-rays dominate, with
  additional, smaller contributions from $^{137}$Cs,
  consistent with simulation predictions.
  
  \vspace*{-2 mm}
  \item To account for environmental sources of
    radioactivity, isotopes such as $^{60}$Co, $^{54}$Mn,
    and $^{135}$Xe were simulated within the air gap
    between the copper end-cap and the NaI(Tl) ACV detector.
    These isotopes originate from residual contamination
    in the surrounding materials or air and contribute
    significantly to the intermediate and high-energy
    regions of the spectrum. The presence of intense
    $\gamma$-lines at 1173.2~keV and 1332.5~keV, each
    reaching $\sim$~0.1~counts~kg$^{-1}$keV$^{-1}$day$^{-1}$,
    together with a flat background below 100~keV,
    makes $^{60}$Co the dominant background contributor
    among these isotopes. In contrast the $\gamma$-peak
    at 834.8~keV from $^{54}$Mn, with a lower rate of
    $\sim$~10$^{-2}$~counts~kg$^{-1}$keV$^{-1}$day$^{-1}$
    in low energy below 100~keV flat region, constitutes a
    manageable yet non-negligible component of the overall
    background. $^{135}$Xe registering a rate
    $\sim$~1.2$\times10^{-2}$~counts~kg$^{-1}$keV$^{-1}$day$^{-1}$
    below 100~keV uniform region confirms itself
    the lowest contributor among environmental sources
    considered. Although contributing to the overall
    background, these sources are effectively controlled and
    suppressed through dedicated shielding strategies and
    careful operational practices, ensuring the integrity
    of the experimental measurements. 

  \vspace*{-2 mm}
\item The differential spectra demonstrate that the difference
  between simulation and measurement is generally negligible,
  except at well-known $\gamma$-peaks where partial
  intensity mismatches are observed. In the high-energy region,
  the difference remains approximately $\sim$~0.1~counts~kg$^{-1}$keV$^{-1}$day$^{-1}$, whereas in the
  low-energy region below 100~keV, negative deviations appear
  due to $\gamma$-lines present only in the simulation. These
  observations highlight the need for minor refinements in
  the simulation input and careful cross-checks of the
  experimental data.

\end{enumerate}

The GEANT4-based background modeling captures the main
features of the measured spectrum, with minor deviations at
specific $\gamma$-lines, and provides insight into the
dominant background contributions within the detector
system. These results validate the precision
of the GEANT4 framework for realistic background
estimation and underscore its importance in optimizing
detector design, shielding, and low-background experimental
techniques. The aim of this work is to characterize the
background and develop
analysis methodologies spanning low to high energies. This
framework serves as a guideline for future research, particularly
targeting the energy range from $\mathcal{O}$(100)~eV up to 500~keV,
while explicitly accounting for reactor-on associated $^{135}$Xe
contributions. The methodologies established here provide a
framework for current and future research, including
re-analysis of existing datasets. Since both reactors at
KSNL are now decommissioned, no further reactor-on data
can be collected. Consequently, the TEXONO experiment is
relocating its setup to the Sanmen Nuclear Power Station,
China, for continued reactor neutrino
studies~\citep{Karmakar::2025}. This work will further
contribute to understanding the background environment of
the forthcoming experimental setup.

\acknowledgments
The authors thank TEXONO collaboration for all the
cooperation for this experiment. The author, Lakhwinder
Singh, acknowledges DST-India and Contract
No. F.30-584/2021 (BSR), UGC-BSR Research Start-Up Grant,
India, for financial support. The authors,
Subhasis Parhi, Lakhwinder Singh and Venktesh Singh,
also acknowledge
DST-FIST and DST-PURSE, New Delhi, for financial
support. The author, Subhasis Parhi, highly
acknowledges UGC,
India for Non-NET Research Fellowship.

\end{document}